\documentclass[journal,twoside,final]{IEEEtran}
\pdfoutput=1  

\usepackage[numbers]{natbib}

\usepackage[dvipsnames]{xcolor}
\usepackage{graphicx}
\usepackage{hyperref}
\makeatletter
\@ifpackageloaded{hyperref}{
	\colorlet{lcolor}{blue!40!black}
	\colorlet{ucolor}{blue!50!cyan!50!black}
	\colorlet{ccolor}{green!40!black}
	\colorlet{fcolor}{red!40!black}
	\usepackage{hypcap}
	\hypersetup{colorlinks=true,linkcolor=lcolor,urlcolor=ucolor,citecolor=ccolor,filecolor=fcolor}
}{
	\providecommand{\href}{2}{#2}
	\providecommand{\orcid}{1}{}
}
\makeatother

\usepackage{amsmath,amssymb,amsfonts,amsthm,mathtools}

\usepackage{enumitem}

\newcommand\copyrighttext{%
\footnotesize 
\textbf{Accepted final version.} Accepted for publication in: IEEE Transactions on Automatic Control, Special Issue on Learning and Control, 2023.\\[0.1\baselineskip]
\textcopyright 2022 IEEE. Personal use of this material is permitted. Permission from IEEE must be obtained for all other uses, in any current or future media, including reprinting/republishing this material for advertising or promotional purposes, creating new collective works, for resale or redistribution to servers or lists, or reuse of any copyrighted component of this work in other works.
DOI: \href{https://doi.org/10.1109/TAC.2022.3200948}{10.1109/TAC.2022.3200948}.
}
\newcommand\copyrightnotice{\IEEEpubid{\parbox{\textwidth}{~\\[3\baselineskip]\copyrighttext}}}

\newtheorem{definition}{Definition}
\newtheorem{lemma}{Lemma}
\newtheorem{proposition}{Proposition}
\newtheorem{thm}{Theorem}
\newtheorem{remark}{Remark}
\newtheorem{assumption}{Assumption}

\newtheorem{problem}{Problem}
\newtheorem{example}{Example}

\newcommand{\VBRT}{\ensuremath{V_{\mathrm{BRT}}}}

\newcommand{\X}{\ensuremath{\mathcal{X}}}  
\newcommand{\Q}{\ensuremath{\mathcal{Q}}}  

\newcommand{\XV}{\ensuremath{{\X_V}}}  
\newcommand{\XU}{\ensuremath{{\X_U}}}  
\newcommand{\XF}{\ensuremath{{\X_F}}}  
\newcommand{\QV}{\ensuremath{{\Q_V}}}  
\newcommand{\QU}{\ensuremath{{\Q_U}}}

\newcommand{\U}{\ensuremath{\mathcal{U}}}
\newcommand{\C}{\ensuremath{\mathcal{C}}}
\newcommand{\D}{\ensuremath{\mathcal{D}}}

\newcommand{\UU}{\ensuremath{\mathfrak{U}}}
\newcommand{\UUsafe}{\ensuremath{\mathfrak{U}_{\mathrm{safe}}}}

\newcommand{\R}{\ensuremath{\mathbb{R}}}
\newcommand{\Rp}{\ensuremath{\R^+}}
\newcommand{\N}{\ensuremath{\mathbb{N}}}
\newcommand{\T}{\ensuremath{\mathbb{T}}}

\newcommand{\pstar}{\ensuremath{p^\star}}
\newcommand{\flow}[2]{\ensuremath{\phi_{#1}^{#2}}}

\newcommand{\Tf}{\ensuremath{{T_f}}}

\newcommand{\RsupQU}{\ensuremath{{R}_{\QU}}}
\newcommand{\RsupXU}{\ensuremath{{R}_{\XU}}}
\newcommand{\RinfQV}{\ensuremath{{R}_{\QV}}}
\newcommand{\timecontroller}[2]{\ensuremath{\hat{#1}_{#2}}}
\newcommand{\alphainf}{\ensuremath{\alpha_{\mathrm{inf}}}}
\newcommand{\alphasup}{\ensuremath{\alpha_{\mathrm{sup}}}}

\newcommand{\cost}{\ensuremath{\rho}}  
\newcommand{\costthreshold}{\ensuremath{\eta}}

\newcommand{\sink}{\ensuremath{\sigma}}

\newcommand{\expp}[1]{e^{#1}}
\newcommand{\dirac}[1]{\ensuremath{\delta_{#1}}}

\newcommand{\finitedt}{\ensuremath{{\delta t}}}

\newcommand{\eqdef}{\dot{=}}

\DeclareMathOperator*{\argmax}{argmax}
\DeclareMathOperator{\expected}{\mathbb{E}}

\newcommand{\mailto}[1]{\href{mailto:#1}{#1}}

\newcommand{\steve}[1]{{\color{ForestGreen}Steve: #1}}
\newcommand{\fs}[1]{{\color{blue}Friedrich: #1}}
\newcommand{\pfm}[1]{{\color{red}Pierre-François: #1}}

\newcommand{\todo}[1]{{\color{orange}TODO: #1}}
\newcommand{\needref}{{\color{orange}[REF]}}
\newcommand{\review}[1]{{\color{Plum}\textbf{Reviewer:} #1}}

\newcommand{\eg}{e\/.\/g\/.,\/~}%
\newcommand{\cf}{cf\/.\/~}%
\usepackage[normalem]{ulem}  
\usepackage{marginnote}

\newcommand{\shortenremove}[1]{{\color{Plum}\sout{#1}}}

\renewcommand{\todo}[1]{}
\renewcommand{\needref}[1]{}
\renewcommand{\review}[1]{}
\renewcommand{\steve}[1]{}
\renewcommand{\fs}[1]{}
\renewcommand{\pfm}[1]{}

\renewcommand{\shortenremove}[1]{}
\providecommand{\QEDclosed}{\IEEEQED}
\newcommand{\orcid}[1]{\href{https://orcid.org/#1}{\raisebox{0.2ex}{\,\includegraphics[height=1.75ex]{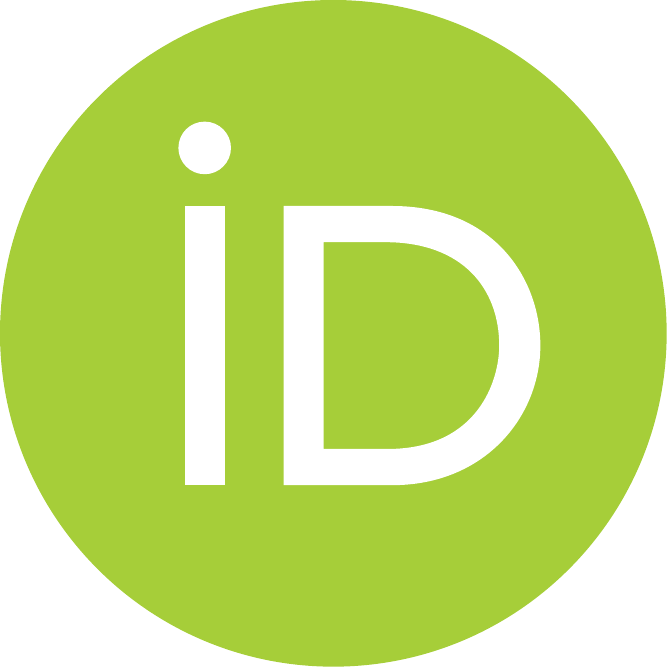}}}}

\begin{document}
    \urlstyle{same}

    \title{Safe Value Functions}
    \author{Pierre-François Massiani\,\orcid{0000-0002-8019-4401}, Steve Heim\,\orcid{0000-0002-4916-7464}, Friedrich Solowjow\,\orcid{0000-0003-2623-5652}, and Sebastian Trimpe\,\orcid{0000-0002-2785-2487}
    \thanks{This work was supported in part by the Cyber Valley Initiative and the
	International Max Planck Research School for Intelligent Systems.}
	\thanks{P.-F. Massiani (\mailto{massiani@dsme.rwth-aachen.de}), F. Solowjow (\mailto{solowjow@dsme.rwth-aachen.de}), and S. Trimpe (\mailto{trimpe@dsme.rwth-aachen.de}) are with the Institute for Data Science in Mechanical Engineering, RWTH Aachen University, 52068 Aachen, Germany.}
	\thanks{P.-F. Massiani, S. Heim (\mailto{heim.steve@gmail.com}), F. Solowjow, and S. Trimpe are with the Intelligent Control Systems Group, Max Planck Institute for Intelligent Systems, 70569 Stuttgart, Germany.}
	\thanks{S. Heim is with the Biomimetic Robotics Lab, Department of Mechanical Engineering, Massachusetts Institute of Technology, 02139 Cambridge, USA.}
    }
    \markboth{Massiani \MakeLowercase{\textit{et al.}}: Safe Value Functions}{Massiani \MakeLowercase{\textit{et al.}}: Safe Value Functions}
    \copyrightnotice
    
    \maketitle
    \begin{abstract}
Safety constraints and optimality are important but sometimes conflicting criteria for controllers.
Although these criteria are often solved separately with different tools to maintain formal guarantees, it is also common practice in reinforcement learning to simply modify reward functions by penalizing failures, with the penalty treated as a mere heuristic.
We rigorously examine the relationship of both safety and optimality to penalties, and formalize sufficient conditions for \emph{safe value functions (SVFs)}: value functions that are both optimal for a given task, and enforce safety constraints.
We reveal this structure by examining when rewards preserve viability under optimal control, and show that there always exists a finite penalty that induces a safe value function.
This penalty is not unique, but upper-unbounded: larger penalties do not harm optimality.
Although it is often not possible to compute the minimum required penalty, we reveal clear structure of how the penalty, rewards, discount factor, and dynamics interact.
This insight suggests practical, theory-guided heuristics to design reward functions for control problems where safety is important.

\end{abstract}

	\IEEEpeerreviewmaketitle

    \section{Introduction}
\IEEEPARstart{V}{alue} functions concisely represent the global behavior of a dynamical system in a scalar function, and have been well studied in both classical control theory and reinforcement learning (RL).
Before solving for the value function, a desired task needs to be accurately translated into a reward function and discount factor.
Often, the main challenge of an engineer is to design a reward that approximates the desired task, but is also dense throughout the state space.
\par
Safety often also needs to be encoded, as many tasks implicitly require that a set of failure states is never visited.
This is particularly challenging, as there may be states that are not failures themselves, but are still unsafe: all trajectories starting from such states end in failure.
This requirement is most naturally written mathematically as a state-space constraint, which can be directly enforced along entire trajectories, such as in model-predictive control~\cite{hewing2019mpcreview}.
It is less clear how the notion of safety can be directly integrated into value functions through rewards.
A common approach is to simply put a large \emph{penalty} on failing.
However, it is still unclear under what conditions this penalty ensures that the resulting value function enforces safety, and when it is a mere heuristic~\cite{garcia_comprehensive_2015, brunke2021safe}.
Furthermore, as adding a penalty is essentially a form of reward shaping, it is unclear whether it affects the optimality of the value function with respect to the original task.\par
We examine the problem from the perspective of viability theory~\cite{aubin2011viability}: a state is in the \emph{viability kernel} if there exists at least one evolution of the system that never fails.
Our core contribution is to examine when rewards preserve viability under optimal control.
We frame this as a condition on the value function, thus exposing the structure of \emph{safe value function} (SVFs): a value function that is both optimal for the original (constrained) task, and avoids failure whenever possible.
We then examine how to obtain SVFs by penalizing failure, and show that there always exists a finite penalty that induces an SVF.
Furthermore, larger penalties do not lose this property.
This insight allows for qualitative predictions on the effects of reward shaping, and provides theory-backed heuristics for designing reward functions when perfect system information is unavailable.
We also examine how SVFs relate to existing concepts and practices, such as strong duality.
This result is of interest for any approach that employs penalties to discourage constraint violation in dynamical systems.
\IEEEpubidadjcol  
\subsection{Related Work}  \label{pos:related work}
Safety in dynamical systems requires constraints to be enforced along entire trajectories, making path planning and tracking a natural approach~\cite{schulman2014SQP}.
However, this approach is often too computationally demanding to be run online for complex systems.
To ease computational burden, constraints are often only enforced on trajectories of finite horizon, such as in sequential trajectory planning~\cite{Holmes2020reachable, shih2020safesets} and model-predictive control (MPC)~\cite{hewing2019mpcreview, nubert2020safeMPC}.
Since these methods only enforce constraints along partial trajectories, guaranteeing safety requires additional conditions to be met.
For example, in MPC, it is known that the choice of terminal constraint plays an important role in guaranteeing safety~\cite[Section\,2.2]{hewing2019mpcreview}. 
To forego this terminal constraint typically requires assumptions of local controllability towards equilibrium~\cite{boccia_stability_2014},~\cite[Chapter\,6]{grune_nonlinear_2017}, which implies viability, to guarantee feasibility and avoid trivially infinite costs.
The latter has also been addressed by introducing discounting~\cite{postoyan_stability_2016}, but viability is still assumed for feasibility.
In contrast, we do not \emph{assume} viability, but study how costs/rewards can be used to guarantee the whole viability kernel's invariance under optimal control.
\par
Trajectory evaluations can be sidestepped by constraining the system to remain in a controlled positive invariant set.
However, these sets are often difficult to compute in practice.
One approach is to formulate safety as an ad-hoc optimal control problem, as studied in Hamilton-Jacobi (HJ) reachability analysis~\cite{bansal2017hamilton}.
The resulting value function can then be used to enforce safety using a backup controller~\cite{fisac2018general, akametalu_reachability-based_2014}, and is completely agnostic in the task formulation.
Another closely related approach is control barrier functions (CBFs)~\cite{Ames2019CBFthapp, choi_robust_2021}, which generalizes the backup controller of HJ reachability with sets of safety-preserving control inputs, which can again be used as task-agnostic backup controllers.
This specialized function can be combined with any existing controller to ensure safety.
However, designing a valid CBF is very challenging in practice, and still requires solving an often equally challenging task of designing an optimal controller.
In this article, we present SVFs as the natural combination of safety and optimality in a single object, and examine the conditions and requirements such that guarantees are not lost.
In particular, the guarantees will rely on choosing a high enough penalty, and only require mild assumptions on the system dynamics. 
Because of this, we expect that SVFs will be amenable to popular reinforcement learning algorithms.
\par
Even when accurate models are available, finding reachable sets or CBFs typically requires extensive system knowledge or computational effort~\cite{chen_hamiltonjacobi_2018, herbert_scalable_2021}.
This has motivated the interest in learning or refining these objects from data~\cite{boffi2020certificates, robey2020cbfDemos, heim2019learnable, chou2021ConstraintGrid}.
A novel approach aims at opening up the toolbox of reinforcement learning (RL) for the computation of reachable sets~\cite{fisac2019reachRL, shih2020safesets}.
In the usual case where the optimal behavior is also learned from data, however, a competition arises between exploring for safety and for optimality.
This creates the need for exploration strategies that balance these two separate objectives~\cite{massiani_exploration_2021, turchetta2019safe}.
SVFs are immediately relevant to this branch of research, as they are also amenable to standard reinforcement learning methods.
\par
The idea of accounting for failure directly in the control objective is not new~\cite{geibel_reinforcement_2001}.
A pragmatic way of doing this is by \emph{penalizing} failure states. 
The penalty is then seen as a design choice controlling the risk of optimal controllers~\cite{geibel_risk_sensitive_2005}, or as a parameter that must be optimized over to allow for the least conservative safe behavior~\cite{schulman2014SQP, paternain_safe_2019}.
From a theoretical perspective, it is still unclear whether optimal controllers of penalized problems are optimal with respect to the original constrained task~\cite{garcia_comprehensive_2015, marco2021robot}, or if they even enjoy safety guarantees; recent literature suggest that they only achieve a weak form of safety~\cite[Section\,2.1.3]{brunke2021safe}.
From a practical perspective, the penalty choice is treated as a heuristic; nonetheless, controllers obtained with this approach often yield good results~\cite{meduri2021deepq, yeganegi2019robust, el2020towards}, and are frequently deployed on real system~\cite{funk2021ETC, rai2018bayesian, Levine-RSS-19, hwangbo2019learning}.
Indeed, a recent benchmark~\cite[Section\,5.3]{ray_benchmarking_2019} reports that penalty methods find nontrivial, constraint-satisfying controllers on the safety-gym suite, outperforming constrained RL algorithms such as CPO~\cite{achiam_constrained_2017}.
Our results reconcile this gap between theory and practice.
We prove that there always exists a finite penalty, such that the resulting value function is both safe and optimal. 
Furthermore, larger penalties still enjoy this property.
\par
In practice, policies and value functions are typically parameterized and solved numerically.
In such cases, choosing an arbitrarily large penalty may result in poor exploration~\cite{li21augmented} and/or an ill-conditioned problem.
Such numerical issues are well known in penalty methods used in classical constrained optimization~\cite[Section 5.4]{martins2021engineering}.
Algorithms such as augmented Lagrangian methods address this by iteratively adjusting the penalty, such that it remains small yet sufficiently large to converge to within an arbitrary approximation error~\cite{li21augmented}.
For clarity of exposition, we keep the focus of this article on the mathematical objects, and do not tackle the issue of approximation errors, which is an active field of research~\cite{paternain_safe_2019}.
It is reasonable to expect that, just as in penalty methods, it is desirable to choose the smallest sufficiently large penalty, even though in theory larger penalties still result in SVFs.
The conditions we derive reveal clear structure of how a sufficiently large penalty depends on discounting, properties of the system dynamics, and reward shaping.
Although a closed-form solution is often not attainable, this structure nonetheless provides clear intuition for choosing useful penalties in practice.
\par
Optimizing a return under a running-cost constraint is also studied as constrained Markov decision processes (CMDPs)~\cite{altman_constrained_1999}.
They provide the necessary theory to describe the solutions to problems with bounded-risk constraints.
We explain in detail how our problem relates to general CMDPs and, specifically, to their duality properties.
It is also known that tools for unconstrained MDPs such as value functions are only marginally useful for the analysis of general CMDPs~\cite{altman_constrained_1999, zheng_constrained_2020}.
These concerns are void for our problem and, in fact, our analysis heavily relies on value functions.
Indeed, we show that our problem is essentially an \emph{unconstrained} MDP in a controlled-invariant set (the viability kernel).
\par
Finally, there is substantial work that aims at iteratively solving the constrained optimal control problem while guaranteeing safety of the (potentially still suboptimal) controller at every iteration~\cite{achiam_constrained_2017, moldovan_safe_2012, turchetta2019safe, shih2020safesets}. These approaches typically restrict exploration to enforce safety, and often allow for convergence to a safe but suboptimal (with respect to the original problem statement) controller.
Our work is of independent interest; we do not focus on learning safely, but on the safety and optimality properties of the optimal controller for the unconstrained problem.
\par

\subsection{Outline}
The article is organized as follows.
In Section~\ref{pos:preliminaries}, we begin with preliminaries and notations.
In Section~\ref{pos:problem}, we state the considered problem.
In Section~\ref{pos:strong duality}, we continue with our two main contributions, i) the concept of safe value functions; and ii) how to enforce it with penalties. 
In Section~\ref{pos:connections}, we discuss the connections between SVFs and related approaches such as strong duality in depth.
In Section~\ref{pos:case study}, we illustrate the exposed structure in different settings with simple examples, and interpret the structure to suggest practical heuristics for choosing penalties, thresholds, and reward functions.
In Section~\ref{pos:outlook}, we conclude with a brief summary, discussion, and outlook.
Appendix~\ref{pos:safe optimal controllers proof} contains the proof of our main theorem and Appendix~\ref{pos:examples finite tf} illustrates Remark~\ref{rmk:nonlipschitz dynamics} with examples.
    \section{Preliminaries and Notations}  \label{pos:preliminaries}
This section introduces the necessary mathematical objects to define safety and optimality. In particular, we address dynamical systems, value functions, and viability theory. 

\subsection{Dynamical Systems}  \label{pos:dynamics}
Our core contributions holds in both continuous- and discrete-time; unless otherwise stated, we present both results together. 
For brevity, we consider the time variable $t$ belonging to the set $\T$, defined as $\Rp$ or $\N$ as appropriate.
Integrals on $\T$ are defined with the Lebesgue measure when $\T=\Rp$, and with the counting measure when $\T=\N$.

\subsubsection{Continuous-Time System} In continuous time, we consider the following deterministic dynamical system:
\begin{equation}
    \dot{x}(t) = f(x(t), u(x(t))),
    \label{eq:continuous time system}
\end{equation}
where $t\in\R^+$ is time, $x(t)\in\X\subset\R^n$ is the state, $u(x(t))\in\U\subset\R^m$ is a control input, and $f:\X\times\U\to\R^n$ is the dynamics. The function $u:\X\to\U$ is a \emph{feedback controller}, and is assumed to be measurable.
We denote $\UU$ the set of measurable feedback controllers.
Our results rely heavily on trajectories: we define $\flow{x}{u}:\R^+\to\X$ as the trajectory starting in $x\in\X$ and following the controller $u\in\UU$.
We assume that trajectories are well-defined and continuous for all time, which can be enforced with appropriate assumptions on $f$.

\subsubsection{Discrete-Time System} In discrete-time, we consider the following deterministic dynamical system:
\begin{equation}
    x_{t+1} = f(x_t, u(x_t)),
    \label{eq:discrete time system}
\end{equation}
where $t\in\N$. The discrete-time setting allows for weaker assumptions, such as non-continuous dynamics $f$ or finite sets.
Our results readily generalize when stochastic controllers are allowed: we mainly describe properties of unconstrained optimal control problems, whose optimal controllers can always be chosen to be deterministic. Stochasticity is not considered here for clarity; we will discuss this more in Remark~\ref{rmk:deterministic controllers}.
\subsection{Optimality} \label{pos:optimality}
\subsubsection{Reward and Return} 
We assume a measurable and bounded \emph{reward function} $r:\X\times\U\to\R$. 
For a given controller $u\in\UU$ and initial state $x\in\X$, the return $G$ is:
\begin{equation}
    G(x, u) = \int_\T \expp{-\frac{t}{\tau}}\cdot r(\flow{x}{u}(t),\timecontroller{u}{x}(t)) \, dt,
    \label{eq:return}
\end{equation}
where we use the shorthand $\timecontroller{u}{x}(t) = u(\flow{x}{u}(t))$, and $\tau$ is a positive \emph{discount rate}. 
In discrete time, the \emph{discount factor} $\gamma$ is commonly used instead, with:
\begin{equation}
    \gamma = \expp{-\frac{1}{\tau}}. \label{eq:discount factor}
\end{equation} 
The discount rate quantifies the problem-specific time-horizon: the smaller $\tau$, the less influence rewards far in the future have on $G(x, u)$. 
Delayed rewards can nonetheless overcome the effects of discounting if they are large: the effective time horizon results from these competing effects.
Since the reward is bounded, the return is finite for every $u\in\UU$ and $x\in\X$.

\subsubsection{Optimal Control and Value Function} 
We consider the classical setting of Lagrange optimal control and reinforcement learning: given an initial state distribution on \X, find a controller that maximizes the expected value of the return~\eqref{eq:return}.
Such a controller will be called \emph{optimal}. If the maximization is subject to some safety constraint, we say that the controller is \emph{optimal for the constrained problem}.
\par
A natural object for finding and analysing optimal controllers in RL is the (optimal) value function:
\begin{equation}
    v:~x\in\X\mapsto\sup_{u\in\UU}~G(x, u).\label{eq:value function}
\end{equation}
Finding this value function iteratively from data is the goal of a large branch of reinforcement learning~\cite[Section\,3.6]{sutton2018reinforcement}.
An important property of $v$ is that it satisfies the dynamic programming identity for all $t\in\T$ and $x\in\X$~\cite[Lemma\,I--4.2]{fleming_controlled_2006}:
\begin{equation}
    \begin{split}
        v(x) = \sup_{u\in\UU} & \left[\int_0^t\expp{-\frac{s}{\tau}}\cdot r(\flow{x}{u}(s),\timecontroller{u}{x}(s))ds \right.\\
        & \left.+ \expp{-\frac{t+\finitedt}{\tau}}v(\flow{x}{u}(t+\finitedt))\vphantom{\int_0^t}\right],
    \end{split}
    \label{eq:dynamic programming}
\end{equation}
where $\finitedt$ is 1 if $\T = \N$, and $0$ otherwise.
\subsection{Safety and Viability} \label{pos:safety and viability}
We consider safety as a state-space constraint: there is an absorbing, closed \emph{failure set} $\XF\subset\X$ of states that the system should never visit. Safety is defined as avoiding the failure set. \par
Failure is challenging to cope with, even though it is often straightforward to define.
Indeed, it is insufficient to simply constrain the controller to stay in the complement of the failure set at each step, as this constraint might eventually become infeasible: there may be states that have not failed yet but from which failure can no longer be avoided within finite time.
This is addressed through viability theory~\cite[Chapter\,2]{aubin2011viability}:
\begin{definition}[Viability kernel]
    The \emph{viability kernel} \XV~is the set of states from where failure can be avoided for all time:
    \begin{equation}
        \XV = \left\{x\in\X, \exists u\in\UU, \forall t\in\T, \flow{x}{u}(t)\notin\XF\right\}.
        \label{eq:viability kernel}
    \end{equation}
\end{definition}
We call its complement the unviability kernel, and denote it as $\XU = \X\setminus\XV$.
Trajectories can stay in $\XV$ for an arbitrarily long time, but as soon as they leave it, they fail within finite time.
The viability kernel is therefore the largest safe set.
\begin{remark} \label{rmk:failure infinite time}
By definition, trajectories starting in $\XU$ fail within finite time.
We reconcile this with infinite time-horizon by instantly transitioning the system to an invariant sink state when it reaches the failure set.
We forego introducing this formally, as it adds notational burden with no practical benefit.
\end{remark}

A safe controller can thus be defined as:
\begin{definition}[Safe controller]  \label{def:safe controller}
We say that a controller $u\in\UU$ is \emph{safe for the point $x\in\X$} if the trajectory $\flow{x}{u}$ avoids \XF~for all time:
\begin{equation}
    \forall t\in\T,~\flow{x}{u}(t)\notin\XF.
    \label{eq:safety}
\end{equation}
Additionally, we say that a controller is \emph{safe} if it is safe for all points $x\in\XV$.
We define $\UUsafe(x)$ as the set of controllers that are safe for the point $x\in\X$, and $\UUsafe=\bigcap_{x\in\XV}\UUsafe(x)$ as the set of safe controllers.
\end{definition}
From the definition of $\XV$, it is immediately clear that Equation~\eqref{eq:safety} can be replaced with $\flow{x}{u}(t)\in\XV$ for all $t\in\T$.\par
We will later formulate a constrained optimization problem over the set of safe controllers, and consider dual relaxations of this problem: we therefore describe $\UUsafe(x)$ as a level set.
\begin{lemma}
    For $x\in\X$ and $u\in\UU$, define the \emph{(discounted) risk} $\cost$ as:
    \begin{equation}
        \cost(x, u) = \int_\T\expp{-\frac{t}{\tau}}\cdot\dirac{\XF}(\flow{x}{u}(t))dt,
        \label{eq:cost}
    \end{equation}
    where $\dirac{\XF}$ is the Dirac distribution in continuous time, or the indicator function in discrete time. A controller $u\in\UU$ is safe for $x\in\X$ if, and only if, $\cost(x, u) = 0$.
\end{lemma}
\begin{IEEEproof}
The lemma follows from the definition of $\dirac{\XF}$: 
this integral evaluates to $\expp{-{t_f}/{\tau}}$, where $t_f\in\T\cup\{\infty\}$ is the time-to-failure of $u$ from $x$, infinite if the trajectory never fails.
By Definition~\ref{def:safe controller}, $u$ is safe for $x$ if, and only if, $\flow{x}{u}(t)\notin\XF$ for all $t\in\T$, i.e., if, and only if, $t_f=\infty$. 
This concludes the proof.
\end{IEEEproof}

Although the discount term $\expp{-{t}/{\tau}}$ is not necessary for the lemma to hold, it will be useful later to linearly combine the risk $\cost$ with the return $G$.
    \section{Problem Statement: Safety Through Penalization}  \label{pos:problem}

We consider a standard control or reinforcement learning problem with dynamics $f$ (\cf Section~\ref{pos:dynamics}), failure set \XF, viability kernel~\XV, and return $G$ as defined in~\eqref{eq:return}. Our goal is to find an optimal controller for the safety-constrained problem:
\begin{equation*}
    \sup_{u \in \UUsafe}\expected_{x\sim\mu}\left[G(x, u)\right], \tag{$\mathcal{C}$}
    \label{eq:constrained problem}
\end{equation*}
where $\mu$
is the initial state probability distribution, whose support is included in \XV.
Since solving constrained optimal control problems is hard in general, a common approach is to relax the constraint $u\in\UUsafe$ and penalize reaching the failure set \XF.
This is done by replacing the reward $r$ by $r - p\cdot \dirac{\XF}$, where $p\in\Rp$ is a (constant) penalty chosen by the designer. 
We call this unconstrained problem the \emph{penalized problem}, and define its value function as
\begin{equation}
    V_p:x\in\X\mapsto \sup_{u\in\UU}~G(x, u) - p\cdot\cost(x,u). \tag{$\mathcal{P}$}
    \label{eq:penalized value}
\end{equation}
We assume that the supremum is attained and refer to solutions as optimal controllers of $V_p$.
The question we answer is: 
\begin{problem}  \label{problem}
Under what conditions on $p$ are solutions of the penalized problem also solutions of the constrained problem?
In other words, when
are the solutions of~\eqref{eq:penalized value} both \emph{safe} and \emph{optimal} for~\eqref{eq:constrained problem}?
\end{problem}
    \section{Main Result}  \label{pos:strong duality}

This section presents our main result: any large enough penalty results in safe and optimal controllers.
Since the value function induces optimal controllers via the dynamic programming principle, it indirectly encodes the regions of the state space that will be visited, and those that will be avoided. 
We can thus define safety of optimal controllers as a property of the value function directly;
we call such value functions \emph{safe}.
We first combine this observation with viability theory to identify the properties and sufficient conditions for SVFs.
We then show in detail how we can leverage penalties to manipulate the value function and enforce safety.

\subsection{Safety as a Property of the Value Function}  \label{pos:sufficient condition}
The standard definition of safety is a point-wise statement for a fixed controller (\cf Definition~\ref{def:safe controller}).
For this definition, the trajectory must be unrolled to check if it ever enters the failure set.
We generalize this point-wise definition to a uniform notion of safety over optimal controllers.
These controllers are directly induced by the value function;
therefore their safety can naturally be stated as a condition on the value function:

\begin{definition}[Safe value function] \label{def:svf}
Consider a value function $v$ as given in~\eqref{eq:value function}. We say that $v$ is a \emph{safe value function} if all of its optimal controllers are both safe and optimal with respect to~\eqref{eq:constrained problem}.
\end{definition}
We emphasize that all SVFs are value functions in the classical RL sense.
\par
\begin{proposition}
Consider $V_p$ as given in~\eqref{eq:penalized value}.
If $V_p$ is an SVF, then:
\begin{equation}
        V_p(x) = V(x),~\forall x \in \XV,
        \label{eq:optimality on xv}
    \end{equation}
    where $V$ is the \emph{constrained value function}:
    \begin{equation}
        V:\,x\in\XV\mapsto\sup_{u\in\UUsafe}G(x, u).
        \label{eq:constrained value}
    \end{equation}
    \label{prop:optimal constrained controllers}
\end{proposition}
\begin{IEEEproof}
This directly follows from Definition~\ref{def:svf} and the fact that $\cost(x, u) = 0$ for $x\in\XV$ and $u\in\UUsafe$. Since optimal controllers of $V_p$ are in \UUsafe, they also maximize $G(x, u)$ over \UUsafe, which is the definition of $V$.
\end{IEEEproof}
Proposition~\ref{prop:optimal constrained controllers} illustrates how, in an SVF, optimality \emph{is decoupled} from safety.
Crucially, optimal trajectories starting from initial conditions inside the viability kernel $\XV$ remain viable and, therefore, the values in $\XV$ are unaffected by the values outside of it; here, $V_{p|\XV}$ does not depend on $p$ if $V_p$ is an SVF.
This prompts the more general observation that, as long as the reward \emph{in} $\XV$ is the same as in~\eqref{eq:constrained problem}, then all SVFs have the same value in $\XV$.
It follows that modifications of the reward that only affect the value of $\XU$ can be used as a degree of freedom to ensure safety, without affecting optimality.
This can be difficult to do in practice, as the viability kernel itself is rarely known with accuracy.
We will show sufficient conditions for failure penalties to stop affecting values in $\XV$ and only affect $\XU$, resulting in an SVF.
To do so, we first derive a more practical condition to check that a value function is safe.

\subsubsection{The Zeroth-Order Condition for Safety}
Our key intuition is that, for an optimal controller to be safe, it is sufficient that states outside the viability kernel $\XV$ are less attractive than states inside $\XV$.
We first introduce the notations
\begin{equation*}
    \RsupQU = \sup_{\QU}\,r
    ,\quad\text{and}\quad
    \RinfQV = \inf_{\QV}\,r,
\end{equation*}
as well as $\QV=f^{-1}(\XV)\subset\XV\times\U$ and $\QU = (\X\times\U)\setminus\QV$ for $\T=\N$.
Intuitively, $\QV$ is the set of state-control pairs that transition into the viability kernel~\cite{heim2019beyond}.
\begin{thm}[Zeroth-order condition] \label{thm:safe optimal controllers}
    Consider $V_p$ as given in~\eqref{eq:penalized value}. Assume that the following \emph{zeroth-order condition} holds:
    \addtocounter{equation}{-1}
    \begin{subequations}
            \makeatletter
            \def\@currentlabel{$\mathcal{Z}$}
            \makeatother
            \renewcommand{\theequation}{$\mathcal{Z}$\alph{equation}}
            \label{eq:safety inducing}
            \begin{align*}
            \sup_{\XU}V_p~&<~\inf_{\XV}V,&\text{if}~\T=\Rp, 
            \tag{$\mathcal{Z}$a}
            \label{eq:safety inducing:continuous}\\
            \RsupQU + \gamma\cdot\sup_{\XU}V_p~&<~\RinfQV + \gamma\cdot\inf_{\XV}V,&\text{if}~\T=\N.\hphantom{^+} 
            \tag{$\mathcal{Z}$b}
            \label{eq:safety inducing:discrete}
            \end{align*}
        \end{subequations}
    Then, the value function $V_p$ is safe.
\end{thm}
\begin{IEEEproof}
See Appendix~\ref{pos:safe optimal controllers proof}.
\end{IEEEproof} 

Condition~\eqref{eq:safety inducing} ensures that an optimal controller never chooses to leave the viability kernel by ensuring that unsafe actions are immediately sub-optimal.
The ``zeroth-order'' terminology comes from the fact that it directly compares the value of $V_p$ in different states.

\subsection{The Penalty Should Be Big Enough}
The intuition for penalizing the failure set $\XF$ is that it reduces the values along trajectories that end in failure.
A natural question is then if this strategy leads to an SVF.
We will now show that, under an additional assumption, it does: increasing penalties causes $V_p$ to decrease arbitrarily and monotonically in $\XU$.
From Proposition~\ref{prop:optimal constrained controllers}, this also ensures that $V_p$ converges to $V$ inside $\XV$.
It is, therefore, always possible to enforce the zeroth-order condition~\eqref{eq:safety inducing}.
The following assumption makes the above reasoning hold:
\begin{assumption} \label{asmptn:bounded tf}
    The time to failure in $\XU$ is uniformly upper-bounded by $\Tf\in\T$.
\end{assumption}
The time $\Tf$ bounds the finite time-to-failure in $\XU$ given by viability theory.
It can be understood as the time-scale of failure for the system of interest.

\begin{lemma}[Influence of the penalty] \label{lemma:influence of p}
    Under Assumption~\ref{asmptn:bounded tf}, the following holds:
    \begin{enumerate}[label=(\roman*)]
        \item For every $x\in\X$, the function $p\mapsto V_p(x)$ is nonincreasing; \label{stmt:nonincreasing}
        \item The value function is uniformly upper bounded on $\XU$:
        \begin{subequations}
            \begin{align}
                \sup_{\XU}V_p~&\leq~\RsupXU\tau\left(1 - \expp{-\frac{\Tf}{\tau}}\right) - p\cdot\expp{-\frac{\Tf}{\tau}},&\text{if}~\T=\Rp,\label{eq:unviable upper bound:continuous}\\
                \sup_{\XU}V_p~&\leq~\RsupXU\cdot\frac{1 - \gamma^{\Tf+1}}{1-\gamma} - p\cdot\gamma^\Tf,&\text{if}~\T=\N\hphantom{^+},\label{eq:unviable upper bound:discrete}
            \end{align}
            \label{eq:unviable upper bound}
        \end{subequations}
        where $\Tf$ is given in Assumption~\ref{asmptn:bounded tf}, and where we define $\RsupXU$ as:
        \begin{equation}
            \RsupXU = \sup_{\XU\times\U}~r.
        \end{equation}
        \label{stmt:unviable upper bound}
    \end{enumerate}
\end{lemma}
\begin{IEEEproof}
Statement~\ref{stmt:nonincreasing} is an immediate consequence of the fact that the supremum over $\UU$ preserves inequalities.
We prove~\ref{stmt:unviable upper bound} by using classic results on the $\sup$ function:
\begin{equation}
    V_p(x) \leq \sup_{u\in\UU}G(x, u) - p\cdot\inf_{u\in\UU}\cost(x, u),
\end{equation}
for all $x\in\XU$.
The first term of the right-hand side (RHS) of this equation is upper-bounded by: 
\begin{equation}
    \RsupXU\cdot\tau\left(1 - \expp{-\frac{\Tf}{\tau}}\right),
\end{equation}
when $\T = \Rp$, and: 
\begin{equation}
    \RsupXU\cdot\frac{1 - \gamma^{\Tf+1}}{1-\gamma},
\end{equation}
when $\T = \N$. 
Now, we use the fact that the time to failure from $x\in\XU$ is upper bounded by $\Tf$ uniformly in $u\in\UU$ to bound the second term:
\begin{equation}
    \cost(x, u) = \int_0^\Tf\expp{-\frac{t}{\tau}}\dirac{\XF}(\flow{x}{u}(t))dt \geq \expp{-\frac{\Tf}{\tau}}.
\end{equation}
Since this is valid for all $x\in\XU$, this proves~\ref{stmt:unviable upper bound}.
\end{IEEEproof}

The main insight we gain from this lemma is in~\eqref{eq:unviable upper bound}: the supremum of $V_p$ on the unviability kernel can be made arbitrarily low by increasing the penalty. 
Therefore, there is a threshold penalty above which the value function is safe:
\begin{thm}\label{thm:existence pstar}
    Under Assumption~\ref{asmptn:bounded tf}, there exists $\pstar\in\Rp$ such that, for all $p>\pstar$, the penalized value function $V_p$ is safe and the zeroth-order condition~\eqref{eq:safety inducing} holds, with:
    \begin{subequations}
        \begin{align}
            \pstar =& \left[\RsupXU\cdot\tau - \inf_{\XV}V\right]\cdot\expp{\frac{\Tf}{\tau}} - \RsupXU\cdot\tau,\nonumber\\&\quad\text{if}~\T=\Rp,\label{eq:pstar:continuous}\\
            \pstar =& \left[\RsupXU\cdot\frac{1 - \gamma^{\Tf+1}}{1-\gamma} + \frac{\RsupQU - \RinfQV}{\gamma} - \inf_{\XV}V\right]\cdot\frac{1}{\gamma^{\Tf}},\nonumber\\
            &\quad\text{if}~\T=\N\hphantom{^+}.\label{eq:pstar:discrete}
        \end{align}
        \label{eq:pstar}
    \end{subequations}
\end{thm}
\begin{IEEEproof}
This is an immediate consequence of Lemma~\ref{lemma:influence of p}. Indeed, with the value of $\pstar$ given in~\eqref{eq:pstar}, the upper bound~\eqref{eq:unviable upper bound} ensures that~\eqref{eq:safety inducing} holds.
\end{IEEEproof}
As the penalty $p$ increases, the penalized objective $V_p(x)$ decreases. 
For $x \in \XV$, $V_p(x)$ is lower bounded, and after $p$ passes $\pstar$ the value coincides with $V(x)$. 
For $x \notin \XV$, the value $V_p(x)$ goes to $-\infty$ as $p$ goes to $\infty$.\par

Finally, we can answer Problem~\ref{problem} and describe how penalized and constrained problems relate to each other. 
For penalties $p>\pstar$, the two problems are equivalent.
In a nutshell, we provide theoretical justification for penalizing failures based on the following arguments:
\begin{enumerate}[label=(\roman*)]
    \item The critical threshold $\pstar$ is finite;
    \item All maximizers of~\eqref{eq:penalized value} for $p > \pstar$ are safe;
    \item All maximizers of~\eqref{eq:penalized value} for $p > \pstar$ are optimal for the constrained problem \eqref{eq:constrained problem};
    \item The critical threshold $\pstar$ explicitly depends on the maximum time to failure, reward function, and discount rate. \label{stmt:summary dependency}
\end{enumerate}
In practice, \ref{stmt:summary dependency} is critical for scaling the penalty. 
For example, the ratio ${\Tf}/{\tau}$ measures how the time-scale of failure compares to the agent's inner time horizon.
Due to the exponential dependency, the required penalties can be extremely large if the agent is short-sighted, i.e., $\tau$ is much smaller than $\Tf$.

\begin{remark}
    Theorem~\ref{thm:existence pstar} extends to the case where failure is non absorbing by increasing the threshold penalty $\pstar$ accordingly.
    The key idea is to adapt Equation~\eqref{eq:unviable upper bound} to account for the additional reward that can be collected after failure, which is bounded due to discounting.
    For brevity, we leave this derivation to the interested reader.
\end{remark}

\begin{remark} \label{rmk:nonlipschitz dynamics}
The zeroth-order condition~\eqref{eq:safety inducing} is a sufficient condition for a \emph{discontinuous} value function to be safe, and Assumption~\ref{asmptn:bounded tf} is a sufficient condition to enforce this zeroth-order condition.
Both require non-Lipschitz dynamics~\cite[Proposition\,3.1]{bardi_optimal_1997}.
\end{remark}
This is because when the state-control space is continuous and $\T = \Rp$, the dynamics determine whether SVFs are continuous or discontinuous.
To extend the presented results to models with Lipschitz dynamics, higher-order conditions are needed.
Alternatively, one may approximate a Lipschitz model, for example with hybrid dynamics, to account for Assumption~\ref{asmptn:bounded tf} and use the zeroth-order condition.
Indeed, models with Lipschitz dynamics often make predictions of unbounded time-to-failure that do not match empirical observations, even when they are accurate and useful for control.
This mismatch between model predictions and observations lends itself to the common adage ``all models are wrong, some are useful''~\cite{box1976science}; Lipschitz-continuity is a very useful assumption for control, where we typically wish to stay well within the viability kernel, but the same models are typically not accurate for bounding the time-to-failure.
We include in Appendix~\ref{pos:examples finite tf} two examples where Assumption~\ref{asmptn:bounded tf} is reasonable for a system that is otherwise well-modeled with a Lipschitz dynamics model.
    
\section{Connections to Other Approaches for Safe Control}  \label{pos:connections}

We explicitly connect SVFs to Hamilton-Jacobi (HJ) reachability and control barrier functions (CBFs).
In particular, we pinpoint their conceptual differences with SVFs despite apparent similarities.
Further, we put our results into perspective to well-known duality theorems from constrained Markov decision processes (CMDPs).

\subsection{The Viability Kernel as a Level Set}
A common approach to safety is to constrain the system within a known safe set: a set of states which can be made positive invariant through control.
The viability kernel $\XV$ is the largest safe set, and can be recovered from SVFs by simple thresholding under the zeroth-order conditions~\eqref{eq:safety inducing}:

\begin{proposition}[Thresholding]  \label{prop:thresholding}
    Assume $V_p$ is a safe value function such that the zeroth-order condition~\eqref{eq:safety inducing} holds. Let $\alphainf$ and $\alphasup$ be the left-hand side and right-hand side of~\eqref{eq:safety inducing}, respectively, for the corresponding $\T$. For all $\alpha\in(\alphainf,\alphasup]$, we have:
    \begin{subequations}
        \begin{align}
            \XV &= \left\{x,~V_p(x)\geq\alpha\right\},&\text{if}~\T=\Rp, \label{eq:threshold:continuous}\\
            \XV &= \left\{x,~\exists u\in\QV[x], r(x, u) + \gamma V_p(x) \geq \alpha\right\},&\text{if}~\T=\N\hphantom{^+}, \label{eq:threshold:discrete}
        \end{align}
        \label{eq:threshold}
    \end{subequations}
where $\QV[x]\subset\U$ is the slice of $\QV$ in state $x$ and $\QV$ is defined in Theorem~\ref{thm:safe optimal controllers}.
\end{proposition}
\begin{IEEEproof}
    This is an immediate consequence of~\eqref{eq:safety inducing}.
\end{IEEEproof}

\begin{remark}
    The simple expression of \XV~of~\eqref{eq:threshold:continuous} also holds in discrete time with further assumptions on the value function. Assume that:
    \begin{equation}
        \inf_\XV V > \sup_{\XU}V_p. \label{eq:thresholding:additional assumption}
    \end{equation}
    Then, Equation~\eqref{eq:threshold:continuous} holds for discrete time as well with the same values of $\alphainf$ and $\alphasup$ as for continuous time. Note that, under Assumption~\ref{asmptn:bounded tf},~\eqref{eq:thresholding:additional assumption} can always be enforced by further increasing the penalty $p$. A sufficient lower value of $p$ for this to hold can be derived from Lemma~\ref{lemma:influence of p}.
\end{remark}

Expressing safe sets as level sets is at the heart of HJ reachability and CBFs. We explore their connections to SVFs.
\subsubsection{Hamilton-Jacobi Reachability}
The goal of HJ reachability is to find the set of states from where the system can be robustly driven into a target set, which is called the \emph{backwards-rechable tube} of the target set~\cite{bansal2017hamilton}.
For example, the backwards-reachable tube of the failure set is the unviability kernel \XU.
The main method~\cite{fisac2019reachRL,chen_hamiltonjacobi_2018} is to cast this as an optimal control problem and find a value function $\VBRT$ defined similarly as in~\eqref{eq:value function}, but where the integral is replaced with an infimum over time:
\begin{equation}
    \VBRT:x\in\X\mapsto \sup_{u\in\UU}~\inf_{t\geq 0}~r(\flow{x}{u}(t)).
\end{equation}
Here, the function $r$ is the signed distance to the failure set.
Its main difference with the reward used in this article is that $\VBRT$ does not have an integral structure over a trajectory; 
in technical terms, $\VBRT$ is not a Lagrange-type value function~\cite[Section\,III.3]{bardi_optimal_1997}.
It instead captures the \emph{minimum} value of $r$ over the trajectory; $\VBRT(x)$ is the closest distance between the trajectory and the failure set.
Once this value function is known, the backwards-reachable tube (i.e., the unviability kernel) can be recovered as a level set. 
It can then be used to solve a classic optimal control problem with safety guarantees, for instance by applying a backup controller when the system is about to exit the viability kernel~\cite{fisac2018general,chen_hamiltonjacobi_2018}.
\citet{hsu_safety_2021} combine this formulation with a task-related reward, and achieve a value function that is both task-optimal and safe, though not a Lagrange-type one.
\par
For clarity of exposition, we have defined SVFs for Lagrange-type value functions only.
Nonetheless, they extend to other forms of optimal control by changing the cost functional $G$ appropriately in~\eqref{eq:value function} and~\eqref{eq:constrained problem}.
A key advantage of the Lagrange cost~\eqref{eq:return} is its compatibility with classical RL.

\subsubsection{Control Barrier Functions}
To connect SVFs to CBFs, we limit the problem setting to continuous time, $\T = \Rp$.
For details, including other formulations of CBFs, we recommend the review by~\citet{Ames2019CBFthapp}. \par
In a nutshell, a CBF of a closed set $\C\subset\X$ is a continuously differentiable function $h:\D\to\R$, with $\C\subseteq\D$, such that $\C$ is the $0$-superlevel set of $h$.
Additionally, $h$ must satisfy the following two properties: it increases along trajectories in $\D\setminus\C$, and \emph{can locally increase} at the boundary of $\C$. 
This means that for every state at the boundary of $\C$, there exists a control input that makes the derivative of $h$ along the trajectory nonnegative at this point. 
For brevity, we will not introduce a formal definition of CBFs and refer to~\cite[Definition\,2]{Ames2019CBFthapp}. The following proposition connects SVFs to CBFs:
\begin{proposition}  \label{prop:cbf}
    Let $V_p$ be a safe value function such that the zeroth-order condition~\eqref{eq:safety inducing} holds. 
    Assume that $V_{p|\XV}$ is continuously differentiable and $\XV$ is compact.
    Also assume that $\frac{\partial V_{p|\XV}}{\partial x}(x)\neq 0$ for all $x\in\partial \XV$, where $\partial\XV$ is the boundary of $\XV$.
    Then, for all $\alpha\in(\alphainf, \alphasup]$, the function $h = V_{p|\XV} - \alpha$ is a control barrier function of $\XV$. 
\end{proposition}
Here, $\alphainf$ and $\alphasup$ are defined in Proposition~\ref{prop:thresholding}.
This result extends the one presented in~\cite{choi_robust_2021} for value functions coming from HJ reachability analysis. \par
\begin{IEEEproof}
    The fact that $\XV$ is the $0$-superlevel set of $h$ is immediate from Proposition~\ref{prop:thresholding}. 
    We use~\cite[Theorem\,3]{Ames2019CBFthapp}, which states that a differentiable indicator function $g:\mathcal{C}\to\R$ of a compact set $\mathcal{C}$ is a barrier function of this set if i) $\frac{\partial g}{\partial x}(x) \neq 0$ for all $x$ on $\partial\mathcal{C}$; and ii) there exists a control law that renders $\mathcal{C}$ invariant. 
    We take $\mathcal{C}=\XV$ and $g = h$.
    The first condition is assumed, and the second comes from the definition of $\XV$.
\end{IEEEproof}
It is worth noting that CBFs are typically used to enforce positive invariance of a conservative subset of $\XV$ that is specified \emph{a priori}, and not the entire viability kernel $\XV$.
Specifying such a set is often non-trivial, since not all subsets of $\XV$ \emph{can} be made positive invariant.
Indeed, choosing $h = V_{p|\XV} - \alpha$ for $\alpha > \alphasup$ may not yield a CBF unless stronger assumptions are made for the the system dynamics and/or reward function~\cite{heim2019learnable, hsu_safety_2021}.
Alternatively to learning a value-function that is also a CBF, a CBF can also be learned from samples of dynamics transitions alone~\cite{robey2020cbfDemos, korda2020computing}: this is because positive invariance is a property of the system dynamics, and separate from optimality and rewards.

\subsection{Duality Properties of Constrained Markov Decision Processes}

There is a more general set of problems modeled as CMDPs, which deal with maximizing a return while simultaneously constraining a running cost under a given threshold~\cite{altman_constrained_1999}.
Our definition of safety is therefore a special case of CMDPs where the running cost is $\cost$ and the threshold is $0$; we refer to it as the $0$-risk CMDP, as opposed to \emph{bounded-risk} CMDPs.
\par
We will now demonstrate that our results do not generally extend to probabilistic safety with a non-zero threshold.
More specifically, it is not always possible to find a penalty such that \emph{all} solutions of the penalized problem are both feasible and optimal for the bounded-risk CMDP.
This shows that penalties alone are ill-suited to solve RL tasks where a non-zero risk is allowed.
The key argument is that solutions to CMDPs are qualitatively different from the ones of MDPs; optimal controllers are stochastic in general~\cite[Theorem\,3.6\,, Theorem\,3.8]{altman_constrained_1999}.
We illustrate below how this allows the penalized problem to admit solutions that are either unsafe or suboptimal for the bounded-risk CMDP.
We also emphasize why strong duality~--~which is known to hold for CMDPs~\cite[Theorem\,3.6]{altman_constrained_1999},~\cite{paternain_constrained_2019}~--~is not contradictory with this result.
Finally, we pinpoint the special properties of the $0$-risk CMDP that make the above concerns void.\par
The results presented here are reinterpretations of results from~\cite[Chapter\,3]{altman_constrained_1999} and \cite{paternain_constrained_2019} enabled by our previous results and general duality in optimization~\cite[Chapter\,5]{boyd2004convex}.
We restrict ourselves to discrete time and finite state-control sets $\X\times\U$.

\subsubsection{Strong Duality does not Guarantee Safety} \label{pos:duality no guarantees}
The bounded-risk CMDP~\eqref{eq:bounded risk problem} differs from~\eqref{eq:constrained problem} in that the risk should only be lower than some threshold $\costthreshold>0$ in the constraint:
\begin{equation}
    \begin{split}
    \sup_{u \in \UU}&\quad\expected_{x\sim\mu}\left[G(x, u)\right],\\
    \textrm{s.t.}&\quad \cost(x, u) \leq \costthreshold, \quad \forall x\in\XV.
    \end{split}
    \tag{$\mathcal{C}_\costthreshold$}
    \label{eq:bounded risk problem}
\end{equation}
In contrast,~\eqref{eq:constrained problem} is equivalent to~\eqref{eq:bounded risk problem} with $\eta = 0$.
Strong duality is known to hold in general for~\eqref{eq:bounded risk problem}~\cite[Theorem\,3.6]{altman_constrained_1999},~\cite{paternain_constrained_2019}. 
Here, strong duality guarantees the existence of a penalty $\pstar$ such that there is \emph{at least one} controller $u^\star\in\UU$ such that $u^\star$ is feasible and optimal for both the original problem~\eqref{eq:bounded risk problem} and the penalized problem~\eqref{eq:penalized value}~\cite[Section\,5.5.5]{boyd2004convex}.
However, not \emph{all} solutions of the penalized problem~\eqref{eq:penalized value} satisfy the constraints of~\eqref{eq:bounded risk problem}.
In other words, some solutions of the dual\footnote{
    We abuse notation here by not considering the constant term $\pstar\cdot\costthreshold$ in the dual problem~\eqref{eq:penalized value}, which is harmless as we only consider the $\argmax$.
} are not feasible for the primal~\cite[Section\,5.5.5]{boyd2004convex}.
The outcome of an RL algorithm is therefore not guaranteed to be safe when $\costthreshold > 0$, even with the optimal penalty $\pstar$.
In fact, primal-dual algorithms for RL typically \emph{assume} feasibility of the found solutions~\cite[Assumption~2]{tessler_reward_2018}.
We explain below how optimal but unsafe solutions can arise, based on an example and theoretical results from~\cite[Chapter\,3]{altman_constrained_1999}.

\begin{example}[Unsafe Solutions with $0$ Duality Gap] \label{ex:unsafe 0 gap}
    \begin{figure}[h]
        \centering
        \includegraphics[width=\columnwidth]{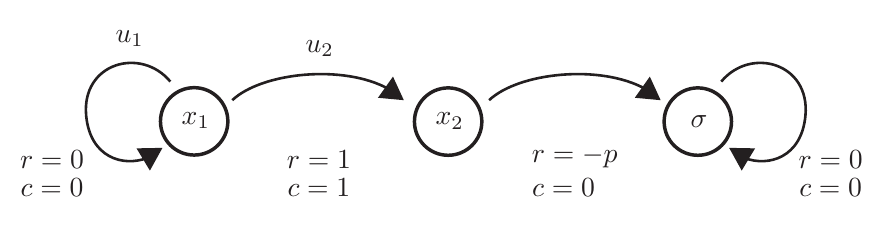}
        \caption{The failure set is $\XF=\{x_2\}$, and the viability kernel $\XV=\{x_1\}$. The system is initialized in $x_1$. Two control inputs are available: $u_1$, which keeps the system in $x_1$, and $u_2$, which transitions it to $x_2$ first and then the sink state $\sink$ (in accordance with Remark~\ref{rmk:failure infinite time}). The reward and cost are both $0$ for $u_1$, and $1$ for $u_2$. The transition to the sink state $\sink$ is penalized with penalty $p\in\Rp$, which is equivalent to a reward of $-p$. Here, the maximum time-to-failure is $\Tf=0$, since the agent immediately lands in $\XF$ after leaving $\XV$.}
        \label{fig:2 state mdp}
    \end{figure}
    Consider the MDP presented in Figure~\ref{fig:2 state mdp}, with $\gamma = 0.1$ and $\costthreshold = 0.01$.
    The corresponding bounded-risk CMDP is:
    \begin{equation}
        \begin{split}
            \argmax_{u}~&\expected\left[G(x_1, u)\right],\\
            \textrm{s.t.}~&\expected\left[\cost(x_1, u)\right] \leq \costthreshold,
        \end{split}\label{eq:cmdp}
    \end{equation}
    where the controller is allowed to be stochastic, thus requiring expected values.
    This problem boils down to finding the optimal probability $\theta^\star$ that should be assigned to the control input $u_2$:
    \begin{equation}
        \begin{split}
            \argmax_{\theta\in[0,1]}~&\gamma\cdot\theta,\\
            \textrm{s.t.}~&\gamma\cdot\theta \leq \costthreshold,
        \end{split}\label{eq:cmdp:original}
    \end{equation}
    which has solution $\theta^\star = 0.1$.
    Now, consider the penalized problem, with $p\in\Rp$:
    \begin{equation}
        \argmax_{\theta\in[0,1]}~\gamma\cdot\theta\cdot(1-\gamma\cdot p).\label{eq:cmdp:penalized}
    \end{equation}
    Strong duality holds with penalty $p=\pstar=10$, such that $(1-\gamma p)=0$; this is the only penalty for which $\theta^\star = 0.1$ also maximizes \eqref{eq:cmdp:penalized}. However, we trivially see that any $\theta\in[0,1]$ has the same objective value for \eqref{eq:cmdp:penalized}.
    \end{example}
    We observe the following:
    \begin{enumerate}[label=(\roman*)]
        \item The set of optimal policies of~\eqref{eq:cmdp:penalized} does not depend on $\costthreshold$ as long as $\eta > 0$;
        \item Not \emph{all} solutions of the penalized problem~\eqref{eq:cmdp:penalized} are solutions of the bounded-risk problem~\eqref{eq:cmdp}; other solutions are either suboptimal or infeasible;
        \item The solution of~\eqref{eq:cmdp} is stochastic.
    \end{enumerate}

Example~\ref{ex:unsafe 0 gap} illustrates the reasons why simply penalizing failure to solve bounded-risk CMDPs may be problematic.
In general, solutions of bounded-risk CMDPs are \emph{stochastic}, contrary to unconstrained MDPs~\cite[Theorems\,3.6\,and\,3.8]{altman_constrained_1999}.
Intuitively, this stochasticity enables a tradeoff between visiting high-cost and high-reward regions ($\XF$ in Example~\ref{ex:unsafe 0 gap}) and low-cost but low-reward ones ($\XV$ in Example~\ref{ex:unsafe 0 gap}).
Unconstrained MDPs can only yield such stochastic controllers when multiple control inputs in a given state have the same value~\cite[Chapter\,3]{sutton2018reinforcement}.
Therefore, the optimal penalty~$\pstar$ is the one that makes the values of safe control inputs equal to the ones of unsafe control inputs; in Example~\ref{ex:unsafe 0 gap}, controls $u_1$ and $u_2$ both have value $0$ for $p = \pstar$.
Consequently, the optimal return does not depend on how optimal controllers choose between such control inputs, and the whole spectrum between being overly conservative and always failing is allowed.
This shows that some solutions of the penalized problem are \emph{suboptimal} for the bounded-risk problem, and that others can be \emph{infeasible}.
For this reason, strong duality alone guarantees neither safety nor optimality of the solutions of the penalized problem when $\costthreshold > 0$. 
In that sense, our result is stronger than strong duality, as we explain in the next section.

\subsubsection{An Unconstrained MDP in the Viability Kernel}
The above analysis stresses the special and beneficial properties of the $0$-risk problem~\eqref{eq:constrained problem}.
It is fundamentally different from general bounded-risk CMDPs~\eqref{eq:bounded risk problem}, with $\costthreshold > 0$.
In a nutshell, the $0$-risk problem reduces to an unconstrained MDP inside of \XV.
\par
The constraint in~\eqref{eq:constrained problem} exhibits a very specific property compared to other bounded-risk constraints, in that it can be directly translated as a \emph{constraint on trajectories}. 
Namely, trajectories should be confined in the viability kernel \XV. 
Therefore, the constrained problem~\eqref{eq:constrained problem} can be reformulated as an unconstrained problem with state space $\XV$ and state-dependent control space $\U(x) = \QV[x]$. 
This interpretation makes very clear why penalization is effective in this case: high enough penalties give low enough values to states in $\XU$, such that optimal controllers are deterministic and stay in $\XV$.

\begin{remark}\label{rmk:deterministic controllers}
Since the $0$-risk problem is essentially an unconstrained MDP in $\XV$, it has deterministic optimal controllers. 
Its optimal controllers also do not depend on the initial state distribution~$\mu$.
This justifies \emph{a posteriori} the fact that we can restrict ourselves to deterministic controllers to solve~\eqref{eq:constrained problem} without loss of generality and why $\mu$ does not play a role in our results.
These properties are not true for general CMDPs~\cite[Chapter\,3]{altman_constrained_1999}; this is why we allowed stochastic controllers in Example~\ref{ex:unsafe 0 gap}.
\end{remark}

    \section{Safe Value Functions and Designing Reward Functions}  \label{pos:case study}

The structure revealed in Equations~\eqref{eq:safety inducing} and~\eqref{eq:pstar} provide important guidelines not just for choosing $\pstar$ and $\alpha$, but for the general problem of designing a reward function.
We first solve a continuous-time example analytically to highlight the relationship of discount rate, time-to-failure, and penalties. 
Then we present a discretized example to illustrate how various forms of reward shaping affect the minimum penalty and thresholds.
The code to reproduce these (and more) examples is publicly available\footnote{\url{https://github.com/sheim/vibly/tree/TAC22/demos/TAC22}}.

\subsection{Continuous-time Example}
To illustrate the influence of the penalty on the value function, we solve the penalized problem analytically on a simple continuous-time example.
We consider the following piecewise-affine system:
\begin{equation} \label{eq:continuous time example}
    \dot{x}(t) = \begin{cases}
        -\frac{L}{\Tf},&\text{if}~x\in(-L, 0),\\
        u(x(t)),&\text{if}~x\in[0, L),\\
        0,&\text{otherwise,}
    \end{cases}
\end{equation}
with state space $\X = [-L, L]$ and finite control space $\U = \{-v, 0, v\}$, where $L$ and $v$ are positive constants.
We define failure with $\XF = \{-L\}$; the trajectory immediately stops when it enters $\XF$.
This results in the viability kernel $\XV = [0, L]$ and unviability kernel $\XU = (-L, 0)$.
It is straightforward to verify that Assumption~\ref{asmptn:bounded tf} holds with constant $\Tf$.
Intuitively, this model can be thought of as a robot driving on a shelf of size $[0, L]$; the robot can move forward or backward at constant speed $\pm v$.
If, however, the robot crosses the ledge of the shelf at $x=0$, it falls at constant speed ${L}/{\Tf}$ until it reaches the ground at $x=-L$ and breaks.

We use the following reward function:
\begin{equation}
    r(x) = \begin{cases}
        -1,&\text{if}~x\in[0, L),\\
        1,&\text{if}~x\in(-L, 0),\\
        0,&\text{if}~x\in\{-L, L\}.
    \end{cases}
\end{equation}
This reward leads the agent to failure from some viable states.
We therefore also penalize failure with the penalty $p$; the total reward is $r - p\cdot\delta_{\XF}$.
We illustrate with this example how a choice of $p > \pstar$ yields an SVF and how $\pstar$, $\tau$, and $\Tf$ interact.

\subsubsection{Scaling the Penalty}
The optimal control consists in either going to and remaining at $x = L$, or falling from the shelf. 
From this observation we deduce the value function for $x\in\XV$:
\begin{equation} \label{eq:penalized value continuous time viable}
    \begin{split}
        V_p(x) = \max~&\left\{\tau \left[\vphantom{e^{-\frac{\Tf}{\tau}}}e^{\frac{x-L}{v\cdot\tau}} - 1\right],\right.\\
        &\left.~~\tau \left[\left(2 - e^{-\frac{\Tf}{\tau}}\right)\cdot e^{-\frac{x}{v\cdot\tau}} - 1\right] \right.\\
        &\left.\qquad- p\cdot e^{-\frac{\Tf + \frac{x}{v}}{\tau}}\right\}.
    \end{split}
\end{equation}
In contrast, in $\XU$, the trajectory goes to failure at speed $\frac{L}{\Tf}$; we get the following expression for $x\in\XU$:
\begin{equation}
    V_p(x) = \tau \left[1 - e^{-\frac{\Tf}{\tau}\left(1 + \frac{x}{L}\right)}\right] - p\cdot e^{-\frac{\Tf}{\tau}\left(1 + \frac{x}{L}\right)}.
\end{equation}
The agent thus remains in $\XV$ for all time if, and only if, the $\max$ in~\eqref{eq:penalized value continuous time viable} is equal to its first argument in $x=0$.
This yields the minimum value for $p$ for an SVF:
\begin{equation} \label{eq:pstar continuous time example}
    p > \tau\left[2 - e^{-\frac{L}{v\cdot\tau}}\right]\cdot e^{\frac{\Tf}{\tau}} - \tau~\eqdef~\pstar.
\end{equation}
The RHS is equal to $\pstar$ as defined in~\eqref{eq:pstar:continuous}; we recover Theorem~\ref{thm:existence pstar} without directly using it.

As $p$ passes $\pstar$, values in $\XU$ keep decreasing and the difference between the thresholds~$\alphainf$ and $\alphasup$ increases (Fig.~\ref{fig:continuous time}, top).
The viability kernel can easily be recovered via thresholding thanks to the discontinuity in $x=0$.
Nevertheless, the value $\alpha = 0$ is not a suitable threshold; this is due to the negative values of the reward in $\XV$. 
We further illustrate this in Section~\ref{pos:design reward}.

\subsubsection{Discount Rate, Time-to-Failure, and Penalties}
\begin{figure}[tbh]
    \centering
    \includegraphics[width=\columnwidth]{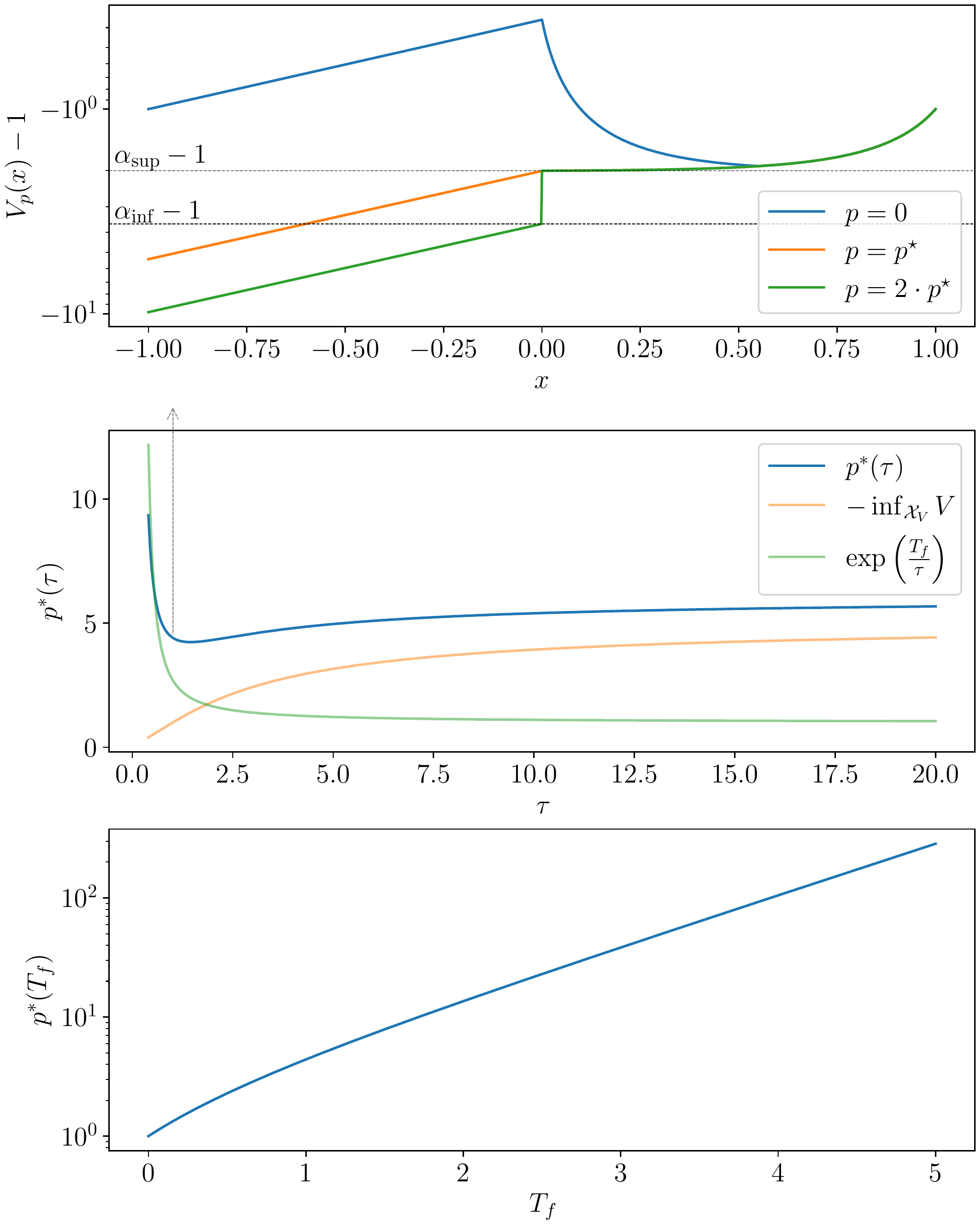}
    \caption{
        (Top) The penalized value function $V_p$ for different values of $p$ with $\tau = 1$. 
        The value on the viability kernel $\XV = [0, 1]$ is not influenced by the penalty when $p \geq \pstar$.
        For $p > \pstar$, the zeroth-order condition holds and $V_p$ is discontinuous.
        (Middle) The dependency of $\pstar$ in the discount rate $\tau$ with $\Tf = 1$.
        For small values of $\tau$, the dependency $e^{\frac{\Tf}{\tau}}$ dominates.
        For large values of $\tau$, the infimum of $V$ on $\XV$ decreases due to the negative rewards in $\XV$; $\pstar$ increases again.
        (Bottom) The exponential dependency of $\pstar$ in $\Tf$, for $\tau = 1$.
        All plots use $L = 1$ and $v = 0.2$.}
    \label{fig:continuous time}
\end{figure}

While Equation~\eqref{eq:pstar} is in general only an upper-bound, it is tight on this example; $\pstar$ is the lowest penalty such that $V_p$ is safe for $p>\pstar$.
This emphasizes the role of the discount rate $\tau$ and time-to-failure~$\Tf$ on the minimum penalty (Fig.~\ref{fig:continuous time}, middle and bottom).
A large $\Tf$ can lead to a large penalty due to the exponential dependency $\pstar\propto e^{{\Tf}/{\tau}}$.
A higher discount rate may help mitigate that problem; it more easily propagates the penalty signal up a failed trajectory.
However, the discount rate is also a crucial parameter determining the safe, optimal behavior.
In fact, if the reward function can be negative inside $\XV$, a large $\tau$ can actually lower $\inf_\XV V$ (the RHS of~\eqref{eq:safety inducing}), increase $\pstar$, and thus encourage failure.

A suitable choice of $\tau$ needs to consider both of these independent criteria, even if the time-to-failure $\Tf$ is the main driver of this choice.
A long $\Tf$ essentially indicates that there exist controllers that can delay an inevitable failure.
While it may seem unintuitive that the existence of such controllers increases the relative gravity of failure, it adds a new interpretation to the benefit of ``fail fast''.
We may conclude that if there are two causally related unviable states, it is advantageous to penalize the one that occurs first.

\subsection{Numerical Example}
To illustrate the influence of the reward on the value function, we use a simple two dimensional system for which we can numerically compute and visualize both the viability kernel and value functions. The system dynamics are
\begin{equation}\label{eq:continuous dynamics numerical example}
    \begin{split}
        \dot{x}_1 & = x_2, \\
        \dot{x}_2 & = - \frac{g}{x_1^2} + \omega^2 x_1 + u,
    \end{split}
\end{equation}
with state $x = \left (x_1, x_2 \right) \in \left[0, 16\right]\times\left[-5, 7\right]$, control input $u \in \left[-1, 1\right]$, and system parameters $g = 10$ and $\omega=0.1$.
The dynamics can be loosely interpreted as a satellite orbiting at constant angular velocity $\omega$, with ${g}/{x_1^2}$ the gravitational force, $\omega^2 x_1$ the centripetal force, and $u$ a radial thruster.
To keep the computation time of value functions low, we discretize the system in space with a 401\texttimes301 grid over the states, and discretize in time by applying the control input $u$ with a zero-order-hold for a period of 1 second, and integrating the continuous time dynamics.
This allows us to explore different combinations of penalties, rewards, and discount factors, and easily visualize the results.
\par
The failure set is $\XF = \{ x: x_1 \leq 1 \text{ or } x_1 \geq 15 \}$, representing the satellite crashing or moving beyond radio communication range.
The continuous-time dynamics~\eqref{eq:continuous dynamics numerical example} don't satisfy Assumption~\ref{asmptn:bounded tf}; nevertheless, the discretized system trivially satisfies it since its state-control space is finite.

All value functions in the following are computed by brute force using value iteration~\cite[Chapter 7.3]{bertsekas2005dynamic}, with a discount factor $\gamma = 0.6$, and $\pstar$ is rounded off to integer values.

\subsection{Success and Failure are Conceptually Independent}
Intuition tells us that success implies avoiding failure, yet avoiding failure does not imply success.
This intuition becomes clear when a task is translated into a parsimonious reward function.
Arguably, the most parsimonious reward describing a task is to simply assign a positive reward for states associated with task success, and zero reward everywhere else.
The definition of a failure set implies that task success should only involve states inside the viability kernel~\XV; the value function for such a reward function is naturally zero outside~\XV, and at least zero inside.
In fact, the value inside $\XV$ is zero only for states from which the task cannot be successfully completed.
We thus see that task success and failure are separate concepts: although the value function (and optimal controller) of a parsimonious reward is automatically safe whenever the task can be successfully completed, it does not enforce the strict inequality condition in~\eqref{eq:safety inducing}.
At the same time, the penalty has no effect on values in $\XV$, and the value function is made safe by any penalty greater than zero.
\par
\begin{figure}[tbh]
    \includegraphics[width=\columnwidth]{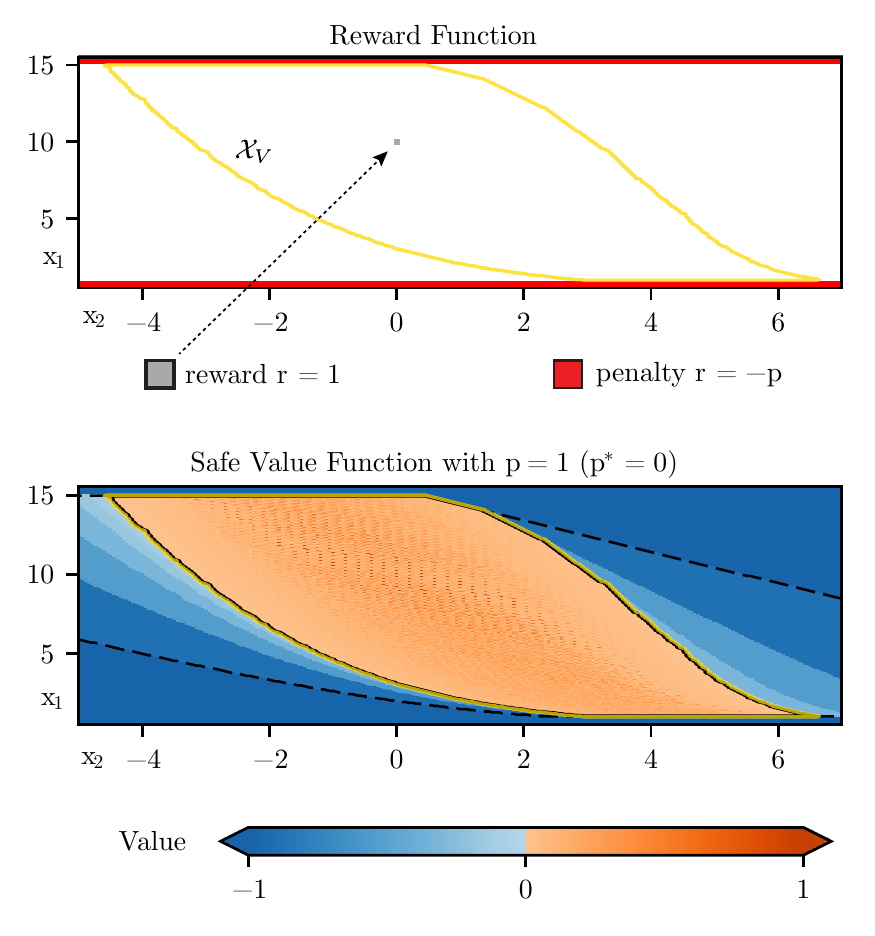}
    \caption{A parsimonious reward function and penalty (top), and the resulting SVF (bottom). The boundary of the viability kernel is outlined in yellow. In this setting, the positive reward is only propagated to its backwards-reachable set, which in this case is the entire viability kernel $\XV$. Conversely, the negative penalty is only propagated along trajectories that cannot avoid failure; that is, to the complement of $\XV$. As such, any penalty will ensure an SVF. The threshold range is $\alpha \in \left[-0.0054, 0\right]$.}
    \label{fig:parsimonious}
\end{figure}

Consider as desired task for our satellite model to reach the equilibrium point at $x=\left(10, 0 \right)$.
A parsimonious reward function is to return one at equilibrium and zero everywhere else.
An SVF for the parsimonious reward function, with penalty $p=1$ assigned to the failure set $\XF = \{ x: x_1 \leq 1 \text{ or } x_1 \geq 15 \}$, is shown in Fig.~\ref{fig:parsimonious}.
Note that for this system, a penalty is not strictly necessary to enforce safety, since the equilibrium point can be reached from any state $x \in \XV$.
\subsection{Recovering the Viability Kernel}\label{ssub:viability}
\begin{figure}[tbh]
    \includegraphics[width=\columnwidth]{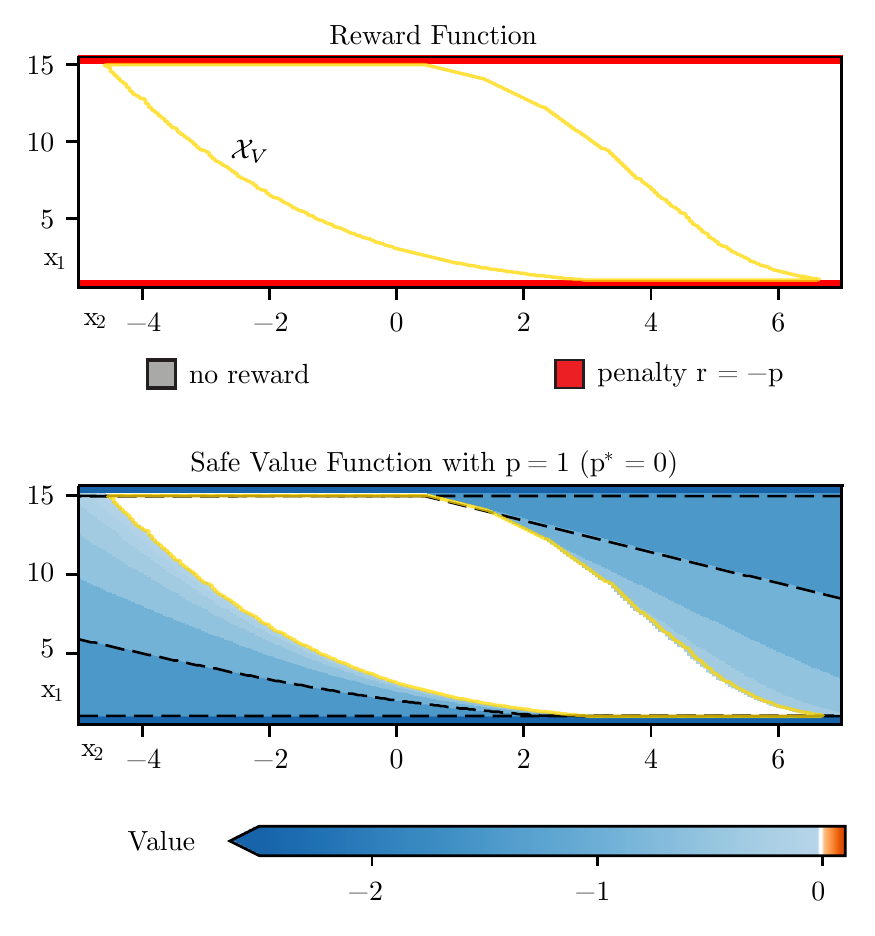}
    \caption{A degenerate reward function and penalty (top), and the resulting SVF (bottom). This setting essentially only considers safety, with no notion of task-related optimality. The resulting SVF is strictly negative outside $\XV$ for any positive value of $p$, and the viability kernel $\XV$ can be easily recovered with the threshold $\alpha = 0$. This setting allows the viability kernel to be recovered with a trivial choice of problem parameters. The threshold range is $\alpha \in \left[-0.0054, 0\right]$.}
    \label{fig:viability}
\end{figure}

The independence of success and failure suggests that we can consider safety independently from any specific task; this property provides the simplest parameters for recovering the viability kernel.
The degenerate reward function $r(x, u) = 0$ does not require even estimates of $\XV$ to design.
It can be immediately seen in Equation~\eqref{eq:pstar} that all terms in $\pstar$ go to zero, and any non-zero penalty will induce an SVF.
We see from the RHS of Equation~\eqref{eq:safety inducing} that $\alphasup$ is also zero, and thus $\XV$ can be recovered by thresholding $V_p$ with $\alpha = 0$, as shown in Fig.~\ref{fig:viability}.

\subsection{Reward Shaping Guidelines}
\label{pos:design reward}
Parsimonious reward functions are rarely used in practice: they are typically too sparse, and it may not be known how to describe task success mathematically.
It is common to design a surrogate reward that is denser, or has a smooth structure more amenable to optimization.
This practice is called reward shaping.
It is when this surrogate reward function is poorly specified that penalties become critical.
\par
We distinguish three categories of modifications, based on whether the reward is positive or negative, and lies inside or outside $\XV$, and examine their effect on $\pstar$ and $\alpha$.
Although $\XV$ is typically unknown \emph{a priori}, over- and under-approximations are sometimes known from domain knowledge~\cite{zaytsev2015two}, and for certain classes of systems, they can be estimated~\cite{korda2020computing, posa2017balancing, manchester2011regions}.
\subsubsection{Benign Rewards}
Adding positive rewards to states in $\XV$ and/or negative rewards in $\X\setminus\XV$ does not increase~$\pstar$ and can, in fact, potentially decrease it.
Such modifications can only increase the RHS and decrease the left-hand side (LHS) of Equation~\eqref{eq:safety inducing}, and thus relaxes the condition for SVFs.
\par
If only benign rewards are used to describe a task, the resulting value function may no longer be optimal compared to a parsimonious reward function, but it remains safe and easy to specify: we have $\pstar = 0$ and $\alpha = 0$.
When combined with other types of rewards, benign rewards reduce $\pstar$.
\subsubsection{Positive Rewards Outside \texorpdfstring{$\XV$}{XV}}
\begin{figure}[bth]
    \includegraphics[width=\columnwidth]{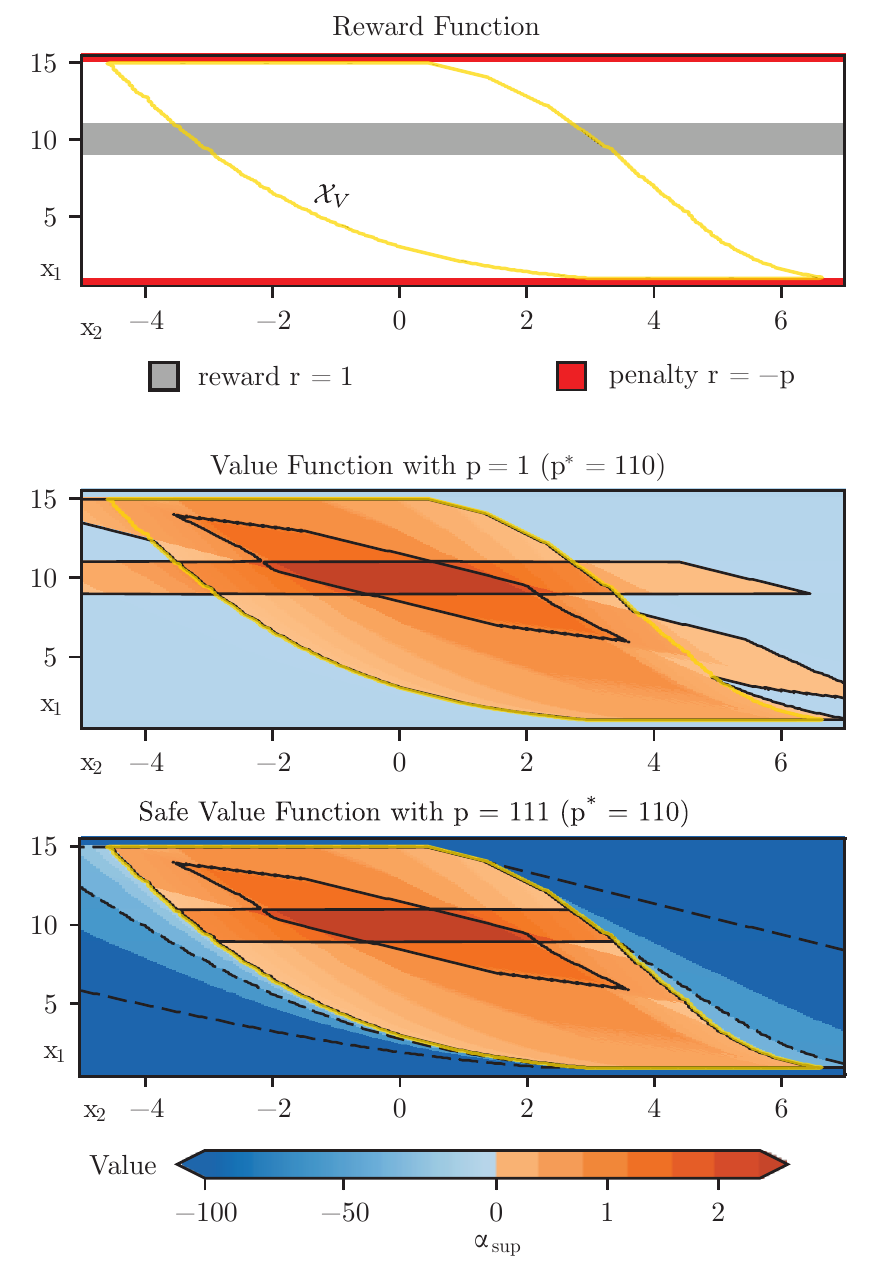}
    \caption{A positive reward function and penalty (top), an insufficiently penalized value function (middle), and an SVF (bottom).
    If a reward function assigns positive rewards outside $\XV$ (top), a larger penalty is required to recover an SVF: with the relatively small penalty of $p=1$, large portions of $\X\setminus \XV$ are already marked with a low value; however, some regions still have a higher value than inside $\XV$. It is only with a penalty of $p=111$, two orders of magnitude greater than the reward, that we recover a value function that is safe everywhere. The threshold range is $\alpha \in \left[-0.004, 0\right]$: we see that, if only positive rewards are used, the viability kernel can still be recovered with a simple threshold at $\alpha=0$.}
    \label{fig:posproxy}
\end{figure}

Assigning positive rewards \emph{outside} $\XV$ can increase $\RsupQU$, which we see in Equation~\eqref{eq:pstar} increases the minimum penalty $\pstar$.
We illustrate this effect with a surrogate reward function that encourages the satellite to remain within a threshold of the equilibrium position, without considering its velocity (Fig.~\ref{fig:posproxy} top):
\begin{equation}\label{eq:posproxy}
r(x,u) = \begin{cases}
        1 & \text{if}, |x_1-10| \leq 1 \\
        0 & \text{otherwise}.
    \end{cases}
\end{equation}

Although the peak of the value function remains centered around the equilibrium point, a relatively high penalty of $p=111$ is needed to recover an SVF (Fig.~\ref{fig:posproxy} bottom). \par
Positive rewards outside $\XV$ can also increase the LHS of Equation~\eqref{eq:safety inducing}, and consequently~\alphainf~in Equation~\eqref{eq:threshold}, making the choice of a thresholding value $\alpha$ more difficult.
Nonetheless, if only positive rewards are used, $\alphasup$ must be non-negative.
There is then always a $p>\pstar$ such that the threshold $\alpha=0$ recovers the viability kernel from an SVF.

\subsubsection{Negative Rewards inside \texorpdfstring{$\XV$}{XV}}
\begin{figure}[bth]
    \includegraphics[width=\columnwidth]{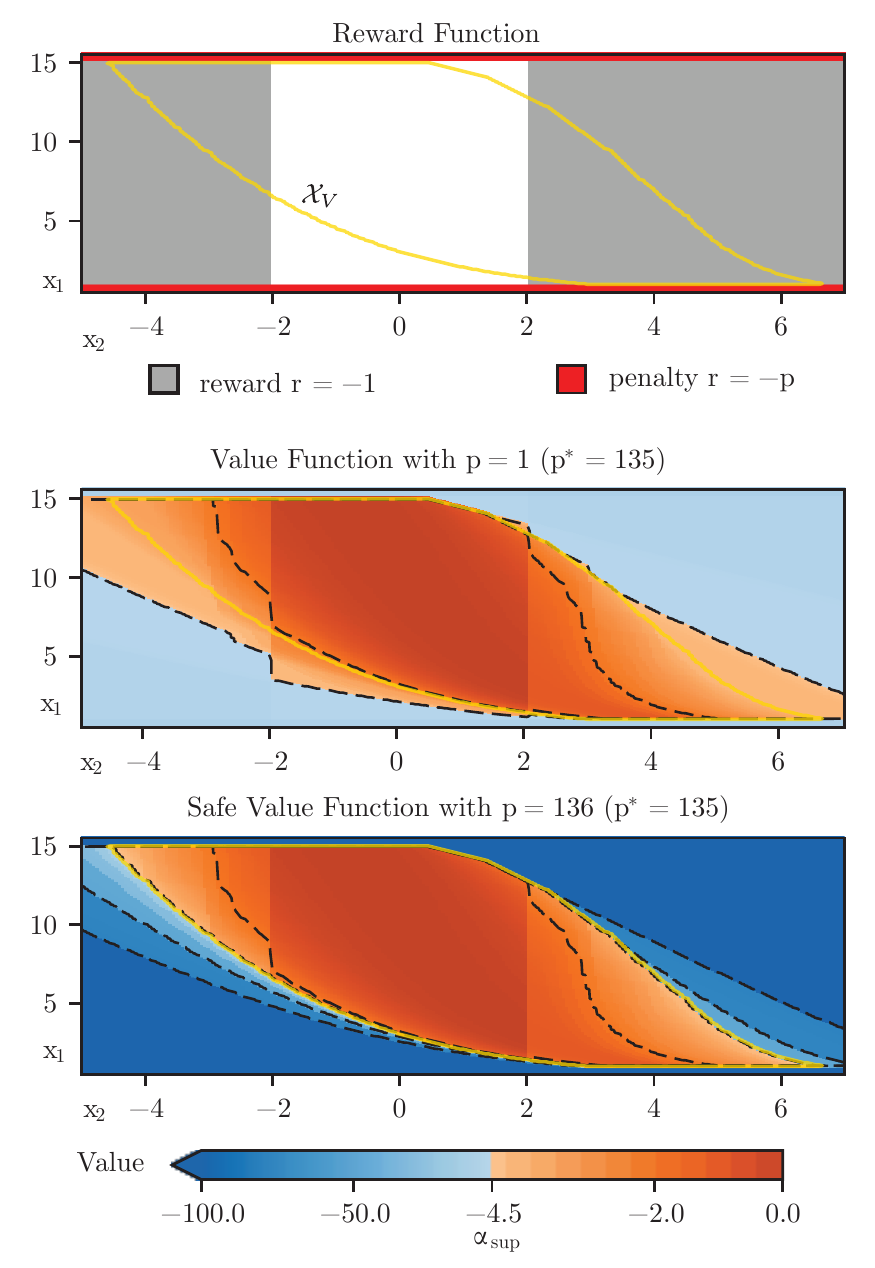}
    \caption{A negative reward function and penalty (top), an insufficiently penalized value function (middle), and an SVF (bottom).
    Note, the quadratic cost on $u$ is not visualized.
    If negative rewards are assigned inside $\XV$, it becomes more difficult to choose not only the penalty $p$, but also the threshold $\alpha$: here the threshold range is $\alpha \in \left[-4.54, -4.53\right]$. 
    We see in the SVF that a choice of $\alpha = 0$ does not recover the viability kernel, but rather a conservative subset. Note that it may not be possible to render these subsets forward invariant; in other words, it is not guaranteed that there exist control barrier functions for these subsets.}
    \label{fig:neg proxy}
\end{figure}

Assigning negative rewards \emph{inside} $\XV$ can decrease $\inf_{\XV}V$ and~\RinfQV, which again leads to a higher minimum penalty~\pstar.
We illustrate this effect with a surrogate reward function that discourages the satellite from high velocities, without considering its position, together with a quadratic cost on the control input (Fig.~\ref{fig:neg proxy} top):
\begin{equation}\label{eq:negproxy}
r(x,u) = \begin{cases}
        -1-u^2 & \text{if}, |x_2| \geq 2 \\
        -u^2 & \text{otherwise}.
    \end{cases}
\end{equation}

Again, the peak of the value function remains centered correctly; however, there are regions of the viability kernel $\XV$ that have a lower value than regions outside, until the penalty is increased to at least $p=136$ (Fig.~\ref{fig:neg proxy} bottom). \par
Furthermore, negative rewards can influence both the RHS and LHS of~\eqref{eq:safety inducing}, making it substantially more difficult to choose a thresholding value $\alpha$.

    \section{Discussion and Outlook}  \label{pos:outlook}
We have identified how rewards can preserve viability through the necessary structure of an SVF: a value function that is both optimal with respect to a given task, and safely avoids a given failure set.
We then provide the zeroth-order sufficient condition for safety, which builds on the intuition that unsafe states should be suboptimal.
Furthermore, we have shown under mild assumptions that the SVF structure can always be enforced by simply shaping the reward function with any sufficiently large penalty on the failure set.
We show how the upper bound for the minimum penalty depends on the reward, discount rate, and system dynamics, and provide an explicit formula (cf. ~\eqref{eq:pstar}) to compute it.
Although the inputs to this formula are often difficult to obtain in practice, they provide insight into the problem, and we discuss several theory-backed heuristics for designing an appropriate reward function.
We expect the insight developed in this article will allow engineers to design reward functions more effectively, while also offering better interpretability of penalty terms.
\par
The new notion of safe value function has many connections to existing tools for safe control, such as Hamilton-Jacobi reachability and control barrier functions.
We examine when these objects are equivalent and when they differ.
Since important objects like the viability kernel can easily be recovered, safe value functions do not lose the advantages of considering safety constraints as separate objects.
We have also shown that penalizing failure is not suited to deal with general bounded-risk constraints.
This result suggests that it is neither necessary nor easier to resort to more elaborate definitions of safety when the dynamics has controllable invariant sets.
\par
The zeroth-order condition we provide is only a sufficient condition, and only applies to systems with discontinuous value functions.
An exciting open question is to derive higher-order conditions for safety of continuous value functions.
This would open the way to other classes of reward shaping for which safety can be checked a priori and thus, facilitate the design of rewards for multiple criteria.
\par
We have restricted our attention to deterministic dynamics. 
Many practical applications account for uncertainty in the model in the form of process noise or exogenous disturbances.
We expect our results to extend to this setting by considering the notion of \emph{invariance kernel}~\cite[Definition\,2.9.2]{aubin2011viability}, sometimes called the discriminating kernel~\cite{fisac2018general}, a robust extension of the viability kernel.
When this invariance kernel is empty, our definition of safety is not suited to describe the system, and one should consider an alternative such as bounded risk constraints.
In this case, we expect that penalizing failures actually harms the optimality of penalized-optimal controllers.
\par
Our long-term goal is to learn provably safe, optimal controllers for the constrained problem.
In practice, value functions and controllers are parameterized to allow for efficient representations in high dimensions, \eg with deep neural networks.
The previous analysis breaks down in this case, as the optimal dynamic programming identity is no longer satisfied and values inside and outside the viability kernel are no longer independent.
Whether a finite penalty ensures safety and constrained-optimality of penalized-optimal, parameterized controllers then remains an important open question.

    \appendix
    
    \subsection{Proof of Theorem~\ref{thm:safe optimal controllers}} \label{pos:safe optimal controllers proof}
We show that any controller that achieves $V_p$ for all $x$ must be safe. 
We start with the following lemma:
\begin{lemma}
    The following conditions are equivalent:
    \begin{enumerate}[label=(\roman*)]
        \item for all $x\in\XV$:
            $$\argmax_{u\in\UU}\,(G-p\cost)(x, u)\subset\UUsafe(x);$$
            \label{stmt:optimal are safe}
        \item for all $x\in\XV$:
            $$V(x) > G(x, u_F) - p\cost(x, u_F),~\forall u_F\in\UU\setminus\UUsafe(x).$$
            \label{stmt:unsafe are suboptimal}
    \end{enumerate}
    \label{lemma:safety cns}
\end{lemma}

\begin{IEEEproof}
To prove the first implication, let $x\in\XV$ and $u_F\in\UU\setminus\UUsafe(x)$. 
We have:
\begin{equation}
    \begin{aligned}
    (G - p\cost)(x, u_F) &< \max_{u\in\UU} (G - p\cost)(x, u),\\
        &= \max_{u\in\UUsafe} G(x, u) = V(x),
    \end{aligned}
    \label{eq:max is constrained value}
\end{equation}
where we used~\ref{stmt:optimal are safe} twice (first to have a strict inequality, and then to restrict the maximization domain to $\UUsafe$) and the fact that $\cost(x, u) = 0$ for all $u\in\UUsafe$.
This proves \ref{stmt:optimal are safe}~$\implies$~\ref{stmt:unsafe are suboptimal}.

We now assume \ref{stmt:unsafe are suboptimal}.
From weak duality:
\begin{equation}
    \max_{u\in\UU}~(G- p\cdot\rho)(x, u) \geq \max_{u\in\UUsafe} G(x, u) = V(x), \label{eq:weak duality}
\end{equation}
because $\rho(x, u) = 0$ for $u\in\UUsafe$.
Now, from \ref{stmt:unsafe are suboptimal}, we have $(G - p\cdot\rho)(x, u_F) < V(x)$ for $u_F\in\UU\setminus\UUsafe(x)$.
Weak duality~\eqref{eq:weak duality} therefore implies that $u_F$ is not in $\argmax_\UU (G - p\cdot\rho)(x, \cdot)$.
This proves that
\begin{equation}
    \argmax_{u\in\UU}~(G - p\cdot\rho)(x, u) \cap (\UU\setminus\UUsafe(x)) = \emptyset
\end{equation}
for all $x\in\XV$, which is equivalent to \ref{stmt:optimal are safe}.
\end{IEEEproof}

Next, we move on to the proof of the main result.
We show that~\eqref{eq:safety inducing} is sufficient for~\ref{stmt:unsafe are suboptimal} to hold. 

Let $x\in\XV$. For $u_F\in\UU\setminus\UUsafe(x)$, consider 
\begin{equation}
    T = \sup\{t\in\T, \flow{x}{u_F}(t) \in\XV\}.  \label{eq:T}
\end{equation}
Such a $T$ is finite, because $u_F$ is unsafe for $x$. 
We now discuss two cases, depending on whether $y=\flow{x}{u_F}(T+\finitedt)$ is in $\XF$.
\par
First, assume that $y\notin\XF$; i.e., $y$ is either in $\XU$ or in $\XV$ (more precisely, on its boundary).
Since $\XF$ is closed, this means that the trajectory spends some time in $\XU$ before reaching $\XF$.
We define a time $\epsilon$ slightly differently depending on whether $\T=\Rp$ or $\T=\N$.
For $\T=\Rp$, we pick $\bar{\epsilon}>0$ such that the trajectory $\flow{x}{u_F}$ stays in $\XU$ until $T+\bar{\epsilon}$.
Next, we simply pick $\epsilon\in(0,\bar{\epsilon})$.
For $\T=\N$, we pick instead $\bar{\epsilon}\geq0$ such that the trajectory stays in $\XU$ until $T + \finitedt + \bar{\epsilon}$ and choose any $\epsilon\in[0,\bar{\epsilon}]\cap\N$. 
This distinction is needed to make $\epsilon$ go to $0$ later. 
The dynamic programming identity for $(G - p\cost)(x, u_F)$ at time $T + \epsilon$ yields:
\begin{equation}
    \begin{split}
        (G - p\cost)(x, u_F) = & \int_0^{T+\epsilon}\expp{-\frac{t}{\tau}}r(\flow{x}{u_F}(t), \timecontroller{u_F}{x}(t))dt \\
            & + \left(G - p\cost\right)(y_\epsilon, u_F)\expp{-\frac{T+\epsilon+\finitedt}{\tau}},
    \end{split}\label{eq:dynamic programming g pcost}
\end{equation}
where we use the shorthand  $y_\epsilon = \flow{x}{u_F}(T+\finitedt+\epsilon)$.
Since $y_\epsilon\in\XU$, we can upper bound $(G-p\cost)(y_\epsilon, u_F)$ by the supremum of $V_p$ on $\XU$:
\begin{equation}
    \begin{split}
        \expp{\frac{T+\epsilon+\finitedt}{\tau}}\cdot (G - p\cost)&(x, u_F) \leq \sup_{\X\setminus\XV}V_p + I_\epsilon \\
        & + \int_{T-\epsilon}^{T+\epsilon}\expp{\frac{T+\epsilon+\finitedt-t}{\tau}}r(\flow{x}{u_F}(t), \timecontroller{u_F}{x}(t))dt,
    \end{split}
    \label{eq:bound unsafe epsilon}
\end{equation}
where we define:
\begin{equation}
    I_\epsilon = \int_{0}^{T-\epsilon}\expp{\frac{T + \finitedt + \epsilon - t}{\tau}}r(\flow{x}{u_F}(t), \timecontroller{u_F}{x}(t))dt.
\end{equation}
Next, we lower bound $V$ by using a safe controller that also collects the return $I_\epsilon$ until time $T-\epsilon$.
We thus construct a safe feedback controller $u$ that follows $u_F$ until $t = T-\epsilon$, and acts safely afterwards. 
Let $b\in\UUsafe$ be any safe backup controller. 
We define $u$ as follows:
\begin{equation}
    u: z\in\X\mapsto\begin{cases}
        u_F(z),~\text{if}~\exists t<T-\epsilon,\,z = \flow{x}{u_F}(t),\\
        b(z),~\text{otherwise}.
    \end{cases}
\end{equation}
This controller is indeed a feedback controller, since it is defined with two feedback controllers on disjoint domains.
It is also safe, since it can be easily checked that $\flow{z}{u}$ never leaves \XV~for any initial state $z\in\XV$.
The dynamic programming equation for $V$ at time $T+\epsilon$ gives:
\begin{equation}
    \begin{split}
        V(x) = \sup_{u_S\in\UUsafe} & \left[\int_0^{T+\epsilon}\expp{-\frac{t}{\tau}}\cdot r(\flow{x}{u_S}(t),\timecontroller{u_S}{x}(t))dt \right.\\
        & \left.+ \expp{-\frac{T+\epsilon+\finitedt}{\tau}}V(\flow{x}{u_S}(T+\epsilon+\finitedt))\vphantom{\int_0^t}\right],
    \end{split}
\end{equation}
We lower bound the supremum by the value in $u\in\UUsafe$. 
By rearranging, we obtain:
\begin{equation}
    \begin{split}
        \expp{\frac{T+\epsilon+\finitedt}{\tau}}\cdot V(x) \geq & \inf_{\XV}V + I_\epsilon \\
        & + \int_{T-\epsilon}^{T+\epsilon}\expp{\frac{T+\epsilon+\finitedt-t}{\tau}}r(\flow{x}{u}(t), \timecontroller{u}{x}(t))dt,
    \end{split}
    \label{eq:bound safe epsilon}
\end{equation}
where we have used $V(\flow{x}{u}(T+\epsilon+\finitedt)) \geq \inf_{\XV}V$ and the fact that $\timecontroller{u}{x} = \timecontroller{u_F}{x}$ for $t$ between $0$ and $T-\epsilon$. \par
We are now ready to conclude. We take $\epsilon\to0$ in~\eqref{eq:bound unsafe epsilon} and~\eqref{eq:bound safe epsilon}. When $\T = \Rp$, this gives:
\begin{align}
    \expp{\frac{T}{\tau}}\cdot V(x) &\geq \inf_{\XV} V + I_0,\label{eq:lower bound v}\\
    \expp{\frac{T}{\tau}}\cdot (G - p\cost)(x) &\leq \sup_{\X\setminus\XV} V_p + I_0,
\end{align}
where we used the fact $\finitedt = 0$.
When $\T=\N$, we have instead $\finitedt = 1$ and use the counting measure for integration, which yields:
\begin{align}
    \expp{\frac{T+1}{\tau}}\cdot V(x) &\geq \inf_{\XV} V + I_0 + \expp{\frac{1}{\tau}}\cdot\RinfQV,\\
    \expp{\frac{T+1}{\tau}}\cdot (G - p\cost)(x) &\leq \sup_{\X\setminus\XV} V_p + I_0 + \expp{\frac{1}{\tau}}\cdot\RsupQU. \label{eq:upper bound g pcost}
\end{align}
For completeness, we should now discuss the case when $y = \flow{x}{u_F}(T+\finitedt)$ is in $\XF$; i.e., when an $\bar{\epsilon}$ cannot be chosen as described after~\eqref{eq:T}. 
The previous computations still hold by simply writing the dynamic programming equations until time $T$ instead of $T+\epsilon$, and defining the controller $u$ with some $\epsilon \in \T$ small enough.
Therefore, Equations~\eqref{eq:lower bound v}--\eqref{eq:upper bound g pcost} hold in this case as well.\par
Since the zeroth-order condition~\eqref{eq:safety inducing} holds, these inequalities imply that
\begin{equation}
    e^{\frac{T+\finitedt}{\tau}}\,V(x) > e^{\frac{T+\finitedt}{\tau}}\cdot(G - p\cdot\rho)(x, u_F).
\end{equation}
As this holds for all $x\in\XV$ and $u_F\in\UU\setminus\UUsafe(x)$, this shows that
\ref{stmt:unsafe are suboptimal} in Lemma~\ref{lemma:safety cns} is satisfied.\hspace*{\fill}~\QEDclosed

\subsection{Finite Time-to-Failure and Lipschitz are conflicting yet useful models}\label{pos:examples finite tf}
Assumption~\ref{asmptn:bounded tf} (the maximum time-to-failure is uniformly upper-bounded) directly conflicts with the Lipschitz-continuity assumption, which is commonly used to model many systems of general interest.
We reconcile this conflict with the observation that these assumptions are useful for separate purposes: Assumption~\ref{asmptn:bounded tf} more closely matches empirical observations when dealing with safety, even when a Lipschitz dynamics model is useful for the purpose of control or prediction. \par
We provide two examples, from the field of legged robotics and from experimental observations of animal motion.
\subsubsection{Linear inverted pendulum model of walking}
A linear inverted pendulum model is commonly used to control foot-step locations of bipedal robots~\cite{khadiv2020walking, koolen2012capturability, zaytsev2015two}. For this model, the relevant dynamics become linear, and it is possible to solve in closed form for the so-called \emph{N-step capture region}~\cite{koolen2012capturability}: the set of foot-step locations from which it is possible to return to a stand-still in $N$ steps.
The formula for the maximum footstep location (which is the border of the capture region) $d_N$ reads~\cite[Eq. 16, rewritten for clarity]{koolen2012capturability}: 
\begin{equation}
d_N = \ell_{max} \sum_{i=1}^{N} (e^{-\Delta t_s'})^i,
\end{equation}
where $\ell_{max}$ is the maximum step-length, and $\Delta t_s'$ is the minimum step-time (time required to swing the foot to the new location). If we take $N$ to infinity, the border of the inifinite-step capture region is 
\begin{equation}
d_\infty = \ell_{max} \frac{e^{-\Delta t_s'}}{1-e^{-\Delta t_s'}}, \label{eq:inft_cap}
\end{equation} and is closed.
Note that, as can be seen in~\cite[Fig. 1]{koolen2012capturability}, the viability kernel is a strict superset of the infinite-step capture region, implying there are states in which it is possible to avoid ever falling down, but also impossible to stop walking. This model also implies that there are states from which it takes infinitely many steps to slow down, and conversely infinitely many steps to fall, even though falling may be unavoidable.
Personal experience tells us this is not true.
In fact, numerical computations for the same and similar models show that if failure can be avoided, it is always possible to come to a stop in just two steps~\cite{zaytsev2015two, zaytsev2018boundaries, heim2019beyond}.
Practice tells us this is true for real robots as well, and experimental observation suggests it is also true for human walking~\cite{matthis2018gaze}.
In other words, this model and its capture regions are very useful for the purposes of control, even though they do not match with real observations on the border $\partial \XV$. This is not a problem because the model's accuracy at the border is not important for control; it does become important 
to estimate time to failure, as in our setting.

\subsubsection{Landing Trajectory of Honeybees}
Our second example is the one developed in~\cite{srinivasan_how_2000}, which studies the time-to-touchdown of honeybees during landing. The authors find that bees essentially follow an exponential trajectory at landing, which implies a theoretically infinite time-to-touchdown. This is explicitly written by the authors~\cite[after Eq. 11]{srinivasan_how_2000}:
\begin{quote}
    ``Thus, the projected time to touchdown does not decrease linearly with time, as it would if the bee approached the surface at constant speed. Neither does the bee decelerate abruptly when the projected time to touchdown falls below a critical value, as flies seem to do before landing on a small target. Rather, the landing bee decelerates continuously, and in such a way as to maintain a constant projected time to touchdown. This, in effect, is the reason why touchdown theoretically requires infinite time.''
\end{quote}
Nevertheless, bees can land. The authors give the following explanation~\cite[after Eq. 4]{srinivasan_how_2000}:
\begin{quote}
    ``This, however, is not a problem in reality because extension of the legs prior to landing ensures that touchdown usually occurs at a nonzero height above the ground.''
\end{quote}
Again, a reasonable model to describe the trajectory of bees yields unrealistic predictions of the ``time-to-reach''.

    \section*{Acknowledgment}
    The authors would like to thank G. Chou, C. Fiedler, J. F. Fisac, K.-C. Hsu, V. Rubies-Royo, and A. von Rohr for insightful discussions, and the anonymous reviewers for their thorough, in-depth comments.
	\bibliography{IEEEabrv,references.bib}
    \vspace{5cm}
    \begin{IEEEbiography}[{\includegraphics[width=1in,height=1.25in,clip,keepaspectratio]{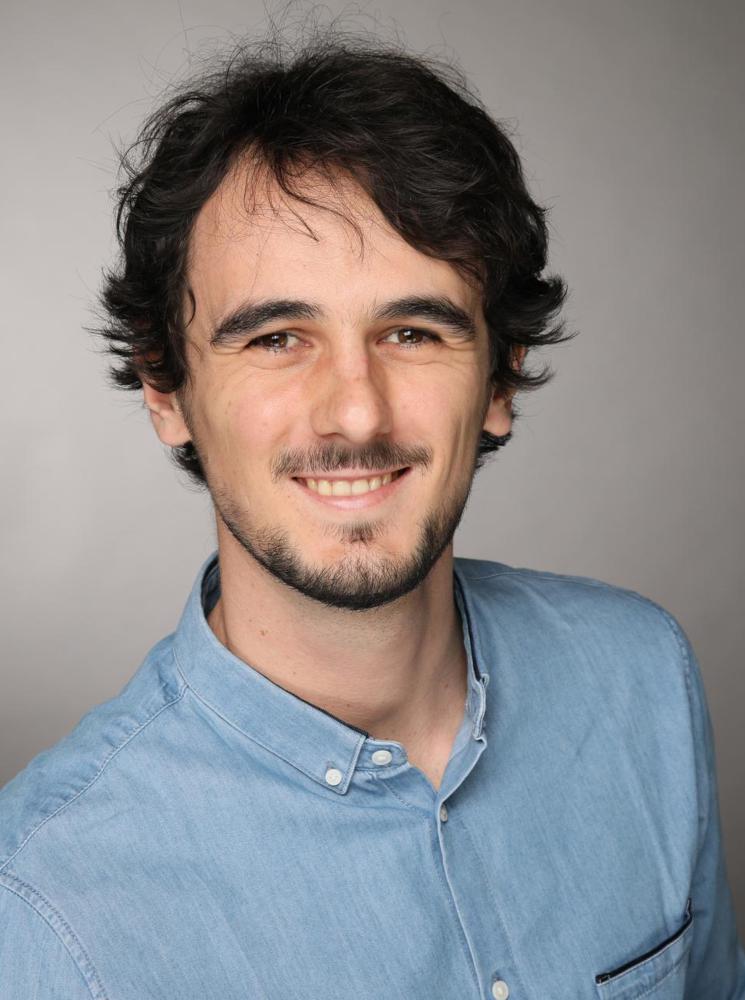}
	}]{Pierre-François Massiani} received the M.Sc. degrees in Science and Executive Engineering from Mines ParisTech, Paris, France, and in Computer Science from PSL University, Paris, France both in 2020. He is currently working toward the Ph.D. degree with the Institute for Data Science in Mechanical Engineering, Aachen, Germany. \par 
	His current research interests involve dynamical systems, control theory, and machine learning.
    \end{IEEEbiography}

    \begin{IEEEbiography}[{\includegraphics[width=1in,height=1.25in,clip,keepaspectratio]{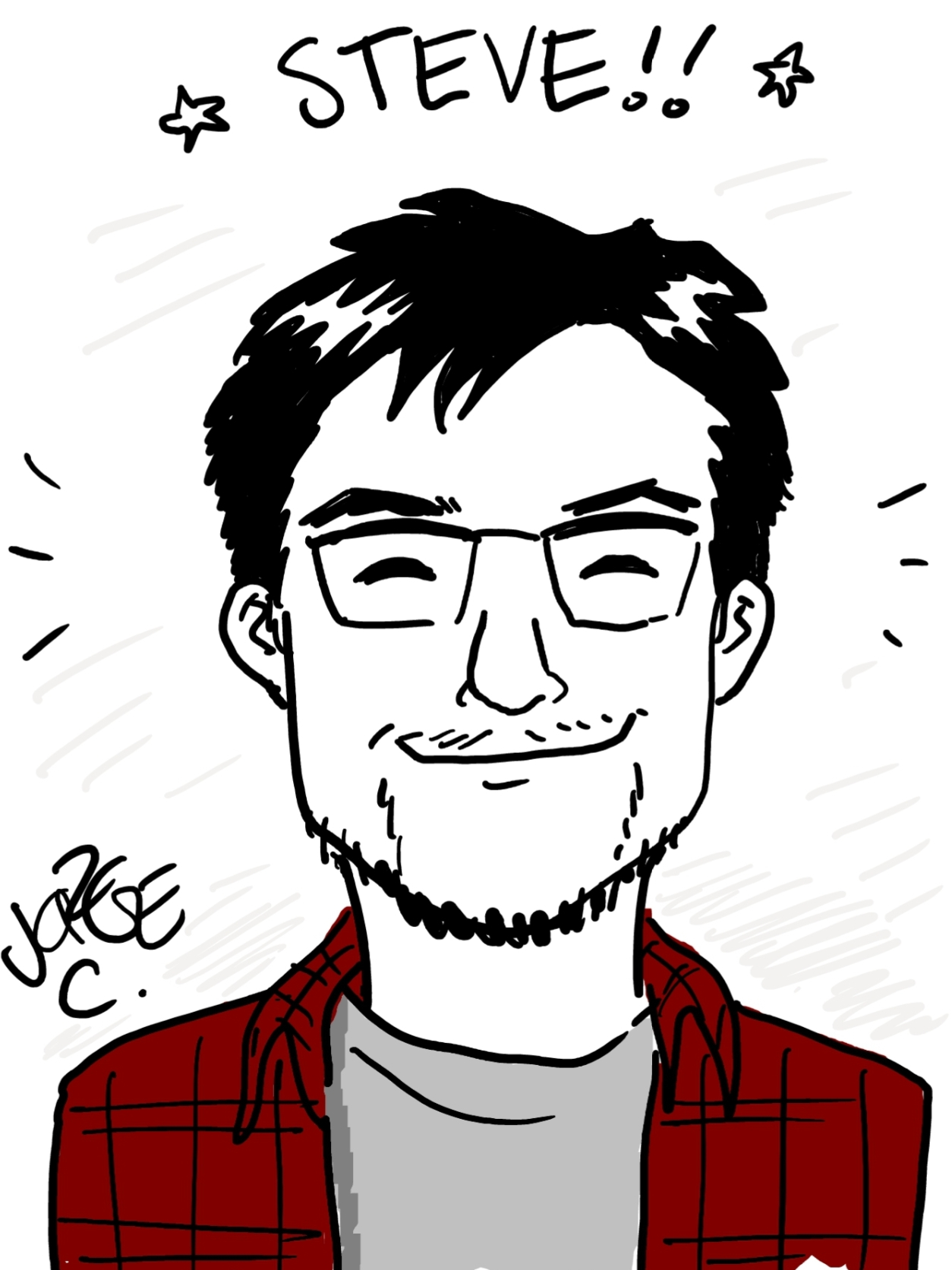}
	}]{Steve Heim} received the B.Sc. degree in mechanical engineering and the M.Sc. in robotics, systems and control from ETH Zurich, Switzerland, in 2012 and 2015, respectively. After a two-year stay at Ishiguro lab at Tohoku University, received the Ph.D. degree from the University of Tübingen through the International Max Planck Research School for Intelligent Systems (MPI-IS) in 2020. \par After a first postdoc with the Intelligent Control Systems Group at MPI-IS, he is now with the Biomimetic Robotics Lab at the MIT. He is interested in dynamics, control, and learning, particularly in relation to movement of animals and machines.
    \end{IEEEbiography}
    
    \begin{IEEEbiography}[{\includegraphics[width=1in,height=1.25in,clip,keepaspectratio]{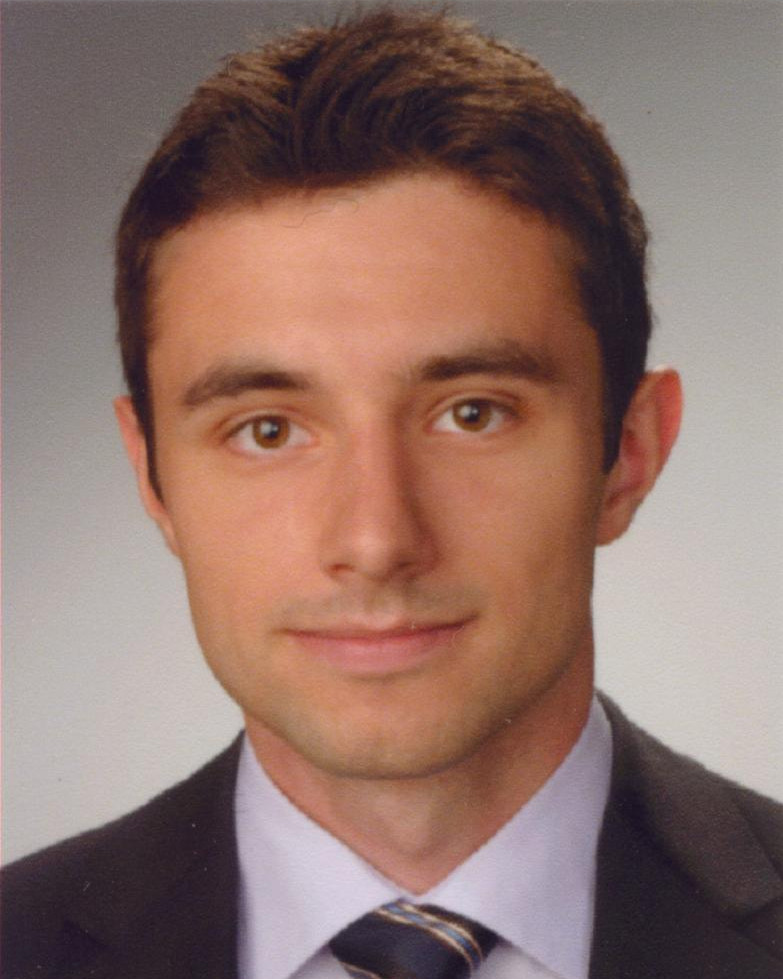}
	}]{Friedrich Solowjow} received the B.Sc. degrees in
    Mathematics and Economics from the University
    of Bonn in 2014 and 2015, respectively, and the M. Sc.
    degree in Mathematics also from the University of
    Bonn in 2017. He is currently working toward the Ph.D. degree with the Intelligent Control Systems Group at the Max
    Planck Institute for Intelligent Systems, Stuttgart,
    Germany. He is a member of the International Max
    Planck Research School for Intelligent Systems. \par
    His main research interests are in control
    theory and machine learning.
    \end{IEEEbiography}

    \begin{IEEEbiography}[{\includegraphics[width=1in,height=1.25in,clip,keepaspectratio]{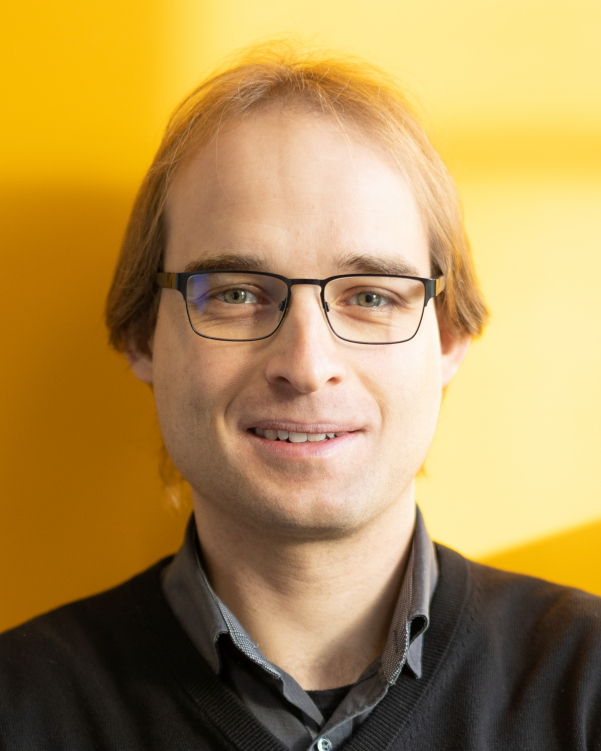}
	}]{Sebastian Trimpe} (M’12) received the B.Sc. degree
    in general engineering science and the M.Sc. degree
    (Dipl.-Ing.) in electrical engineering from Hamburg
    University of Technology, Hamburg, Germany, in
    2005 and 2007, respectively, and the Ph.D. degree
    (Dr. sc.) in mechanical engineering from ETH Zurich,
    Zurich, Switzerland, in 2013.\par Since 2020, he is a
    Full Professor with RWTH Aachen University, Aachen, Germany,
    where he heads the Institute for Data Science in Mechanical Engineering. Before, he was an independent
    Research Group Leader at the Max Planck Institute
    for Intelligent Systems, Stuttgart and Tübingen, Germany. His main research
    interests are in systems and control theory, machine learning, networked and
    autonomous systems. \par
    Dr. Trimpe has received several awards, including the triennial IFAC World Congress Interactive Paper Prize (2011), the Klaus Tschira Award for achievements in public understanding of science (2014), the Best Paper Award of the International Conference on Cyber-Physical Systems (2019), and the Future Prize by the Ewald Marquardt Stiftung for innovations in control engineering (2020).
    \end{IEEEbiography}

\end{document}